\documentstyle[amsfonts,tighten,preprint,eqsecnum,aps,prd]{revtex}
\begin{document}
\draft

\hyphenation{
mani-fold
mani-folds
geo-metry
geo-met-ric
Schwarz-schild
}



\def\BbbR{{\Bbb R}}
\def\BbbZ{{\Bbb Z}}

\def\casehalf{{\case{1}{2}}}

\def\RPthree{\BbbR P^3}
\def\RPtwo{\BbbR P^2}

\def\hilbert{{\sf H}}
\def\hilbertzero{{\sf H}_0}

\def\hzero{H_0}
\def\hhzero{{\hat H}_0}

\def\bm{{\bf m}}
\def\bp{{\bf p}}

\def\bq{{\bf q}}

\def\aspropto{\buildrel \sim \over \propto}

\def\ads{{anti-de~Sitter}}
\def\rnads{{Reissner-Nordstr\"om-anti-de~Sitter}}

\def\lplanck{l_{\rm Planck}}
\def\mplanck{m_{\rm Planck}}


\preprint{\vbox{\baselineskip=12pt
\rightline{DAMTP 96--53}
\rightline{PP96--98}
\rightline{gr-qc/9605058}}}
\title{Area spectrum of the Schwarzschild black hole}
\author{Jorma Louko\footnote{On leave of absence from
Department of Physics, University of Helsinki.
Electronic address:
louko@wam.umd.edu}}
\address{
Department of Physics,
University of Maryland,
College Park,
Maryland 20742--4111,
USA}
\author{Jarmo M\"akel\"a\footnote{Electronic address:
j.m.makela@damtp.cam.ac.uk}}
\address{Department of Applied Mathematics and Theoretical Physics,
University  of Cambridge,
\\
Silver Street, Cambridge CB3 9EW, United Kingdom}
\date{Revised version, September 1996. Published in
{\it Phys.\ Rev.\ \rm D \bf 54} (1996), pp.~4982--4996.
}
\maketitle
\begin{abstract}%
We consider a Hamiltonian theory of spherically symmetric vacuum Einstein
gravity under Kruskal-like boundary conditions in variables
associated with the Einstein-Rosen wormhole throat. The configuration
variable in the reduced classical theory is the radius of the throat, in
a foliation that is frozen at the left hand side infinity but
asymptotically Minkowski at the right hand side infinity, and such that
the proper time at the throat agrees with the right hand side Minkowski
time. The classical Hamiltonian is numerically equal to the
Schwarzschild mass. Within a class of Hamiltonian quantizations, we
show that the spectrum of the Hamiltonian operator is discrete and bounded
below, and can be made positive definite. The large eigenvalues behave
asymptotically as~$\sqrt{2k}$, where $k$ is an integer. The
resulting area spectrum agrees with that proposed by
Bekenstein and others.
Analogous results hold in the presence of a
negative cosmological constant and electric charge. The classical input
that led to the quantum results is discussed.
\end{abstract}

\narrowtext

\section{Introduction}
\label{sec:intro}

One of the most intriguing problems in black hole thermodynamics is the
statistical mechanical interpretation of black hole entropy. One surmises
that black hole entropy should reflect an outside observer's
ignorance about the quantum mechanical microstates of the hole, but it has
proved very difficult to characterize what exactly these quantum mechanical
microstates might be. There are hints that the relevant
degrees of freedom may live on the horizon of the hole
\cite{cartei1,cartei2,carlip,bala-edge}, in situations where the horizon
can be meaningfully defined.
There is also evidence that black hole entropy describes the
entanglement between the degrees of freedom in the interior and in the
exterior of the hole
\cite{bombelli+,srednicki,fro-nov,frolov}.
Recent results from string theory \cite{horo-rev} suggest that black hole
entropy can be recovered by counting the quantum microstates even in
situations where the definition of these states presupposes no black hole
geometry. For reviews, see Refs.\
\cite{horo-rev,page-inf-rev,beken-rev}.

Even prior to Hawking's prediction of black hole radiation
\cite{hawkingCMP}, the anticipated connection between black holes and
thermodynamics led Bekenstein \cite{bekenstein1} to propose that the
horizon area of a black hole is quantized in integer multiples of a
fundamental scale, presumably of the order of the square of the Planck
length
$\lplanck=\sqrt{\hbar G c^{-3}}\,$:
\begin{equation}
A = \alpha k \, \lplanck^2
\ \ ,
\label{area-quantization}
\end{equation}
where $k$ ranges over the positive integers and $\alpha$ is a pure number
of order one.
This proposal has since been revived on various grounds; see Refs.\
\cite{mukha1,kogan1,mazur-grg,mazur-string,%
mukha2,danielsson,bellido,peleg1,maggiore,%
kogan2,lousto2,peleg2,bek-mu,mazur-pol1,mazur-pol2,%
kastrup,barvi-kunst1,barvi-kunst2}, and references therein.
Although the horizon of a (classical) black hole is a nonlocal
object, its total area is completely determined by the irreducible mass
\cite{MTW}, and one can therefore alternatively view the rule
(\ref{area-quantization}) as a proposal for the spectrum of the
quantum irreducible mass operator.\footnote{We thank John Friedman and
Pawel Mazur for emphasizing this to us.} As the irreducible mass can
classically be read off from the asymptotic falloff of the black hole
gravitational field, one expects such an operator to be sensibly definable
even in a quantum theory that only refers to observations made at an
asymptotically flat infinity. In particular, for a Schwarzschild hole, the
irreducible mass coincides with the Schwarzschild mass.

The implications of the area spectrum (\ref{area-quantization}) for
macroscopic physics were recently elaborated on by Bekenstein and Mukhanov
\cite{bek-mu}. Consider, for concreteness, a Schwarzschild hole. The area
is given in terms of the Schwarzschild mass $M$ by
\begin{equation}
A = 16 \pi \left(\lplanck \over \mplanck \right)^2 M^2
\ \ ,
\label{schw-area}
\end{equation}
where $\mplanck = \sqrt{\hbar c G^{-1}}$ is the Planck mass. Now,
(\ref{area-quantization}) and (\ref{schw-area}) imply that $M$ can only take
discrete values. When the black hole evaporates, it can thus only make
transitions between the mass eigenstates corresponding to these discrete
values. As a consequence, the radiation comes out in multiples of a
fundamental frequency, which is of the same order as the maximum of
Hawking's black-body  spectrum, and the corresponding wavelength is of the
order of the Schwarzschild radius of the hole. This means that the
radiation will differ from the black-body spectrum in a way that is, in
principle, macroscopically observable. For example, if $M$ is of the order
of ten solar masses, or $2\times 10^{31}$~kg, then the fundamental
frequency is of the order of $0.1$~kHz, which is roughly the resolving
power of an ordinary portable radio receiver. The discussion can also be
generalized to accommodate a nonvanishing angular momentum
\cite{bekenstein1}.

Arguments presented in favor of the area spectrum
(\ref{area-quantization}) include
quantizing the angular momentum of a
rotating black hole \cite{bekenstein1},
information theoretic considerations
\cite{mukha1,mukha2,danielsson,bek-mu},
string theoretic arguments \cite{kogan1,mazur-grg},
periodicity of Euclidean or Lorentzian time
\cite{kogan1,mazur-string,mazur-pol1,%
kastrup,barvi-kunst1,barvi-kunst2},
a treatment of the event horizon as
a membrane with certain quantum mechanical properties
\cite{maggiore,lousto2},
and a Hamiltonian quantization of a dust collapse
\cite{peleg2}.
Recently, a membrane model for the horizon
\cite{lousto3} recovered an area spectrum that is finer
than~(\ref{area-quantization}), and a calculation within a loop
representation of quantum gravity \cite{barreira} recovered an area spectrum
that effectively reproduces the Planckian spectrum for black hole
radiation. The purpose of the present paper is to give a derivation of
(effectively) the spectrum (\ref{area-quantization}) from a Hamiltonian
quantum theory of spherically symmetric Einstein gravity,
with judiciously chosen dynamical degrees of freedom.

By Birkhoff's theorem \cite{exact-book}, the local properties of
spherically symmetric vacuum Einstein geometries are
completely characterized by a single parameter, the Schwarzschild mass.
In a classical Hamiltonian theory of such spacetimes, the true dynamical
degrees of freedom are thus expected to
contain information only about the Schwarzschild mass and
the embedding of the spacelike hypersurfaces in the spacetime.
It was demonstrated in Refs.\
\cite{thiemann1,thiemann2,kuchar1,LW2} that this is indeed the case,
under certain types of boundary conditions that specify the
(possibly asymptotic) embedding of the ends of the spacelike
hypersurfaces in the spacetime. The variables of the
reduced theory then consist of a single canonical pair: the coordinate
can be taken to be the Schwarzschild mass, and its conjugate momentum
carries the information about the evolution
of the (asymptotic) ends of the spacelike hypersurfaces in the spacetime.
The theory is thus no longer a field theory, but a theory of finitely many
degrees of freedom. This means, in particular, that quantization of the
theory can be addressed within ordinary, finite dimensional quantum
mechanics.

Our Hamiltonian theory of spherically symmetric vacuum spacetimes will be
built on two major assumptions. First, we shall adopt for the spacetime
foliation the boundary conditions of Ref.\ \cite{kuchar1}. This implies
that the classical solutions have a positive value of the Schwarzschild
mass, and that the spacelike hypersurfaces extend on the Kruskal manifold
from the left hand side spacelike infinity to the right hand side
spacelike infinity, crossing the horizons in arbitrary ways. We shall,
however, specialize to the case where the evolution of the
hypersurfaces at the left hand side infinity is frozen, and the evolution
at the right hand side infinity proceeds at unit rate with respect to the
right hand side asymptotic Minkowski time. This means that, apart from
constraints, the Hamiltonian will consist of a contribution from the right
hand side infinity only, and the value of the Hamiltonian is equal to the
Schwarzschild mass. The physical reason for this choice is that
while our theory will remain that of vacuum spacetimes,
we expect these conditions to
correspond to physics accessible to an inertial observer at
one spacelike infinity, at rest with respect to the hole: the proper time
of such an observer is our asymptotic Minkowski time, and the
Arnowitt-Deser-Misner (ADM)
mass observed is the  Schwarzschild mass.

Second, we shall adopt as our reduced dynamical variables a
canonical pair that is intimately related to the dynamical aspects of the
Kruskal manifold. Our configuration variable $a$ can be
envisaged as the radius of the Einstein-Rosen wormhole throat \cite{MTW},
in a spacetime foliation in which the proper time at the throat increases
at the same rate as the asymptotic Minkowski time at the right hand side
infinity. An example of a foliation satisfying these conditions
can be constructed by taking the Novikov coordinates \cite{MTW} and
deforming them near the left hand side infinity to conform to our boundary
conditions there.

The resulting classical theory has two properties whose physical
interest should be emphasized. First, every classical solution is
bounded, in the sense that the variable $a$ starts from zero,
increases to the maximum value~$2M$, where $M$ is the Schwarzschild mass,
and then collapses back to zero. This evolution corresponds to the wormhole
throat starting from the white hole singularity, expanding to the
bifurcation two-sphere, and then collapsing to the black hole singularity.
The spacetime dynamics in these variables is therefore, in a certain
sense, confined to the interior regions of the Kruskal manifold.
This property reflects the physics observed by an inertial
observer at asymptotic infinity, as such an observer sees her
exterior region of the Kruskal manifold as static.
Second, as the proper time on the timelike geodesics that form the
throat trajectory increases at the same rate as the asymptotic
right hand side Minkowski time, one may regard
our foliation as a preferred one, by the principle of equivalence, for
relating the experiences of an inertial observer at the asymptotic
infinity to the experiences of an inertial observer at the throat. Note,
however, that as the total proper time from the initial singularity to the
final singularity along the throat trajectory is finite, our choice of the
foliation implies that the throat reaches the white and black hole
singularities at finite values of the asymptotic right hand side Minkowski
time. As the asymptotic right hand side Minkowski time
evolves at unit rate with respect to our parameter time,
this means that the classical theory is incomplete: the classical solutions
cannot be extended to arbitrarily large values of the parameter time,
neither to the past nor to the future.

We quantize the theory by Hamiltonian methods, treating $a$ as a
configuration variable, specifying a class of `reasonable' inner
products, and promoting the classical Hamiltonian into a self-adjoint
Hamiltonian operator. For certain choices of the inner product the
Hamiltonian operator turns out to be essentially self-adjoint, whereas
in the remaining cases the class of self-adjoint extensions is parametrized
by $U(1)$ and associated with a boundary condition at $a=0$. We find that
the spectrum of the Hamiltonian is discrete and bounded
below in all the cases. When the Hamiltonian is essentially self-adjoint,
the spectrum is strictly positive, and in the remaining cases there always
exist self-adjoint extensions for which the spectrum is strictly positive.
A~WKB estimate for the large eigenvalues of the Hamiltonian yields,
via (\ref{schw-area}), the result that the large area eigenvalues are
asymptotically given by\footnote{We shall use the symbol $\sim$ to denote
an asymptotic expansion throughout the paper.}
\begin{equation}
A \sim 32\pi k \, \lplanck^2 + \hbox{constant} + o(1)
\ \ ,
\label{area-asymptotic}
\end{equation}
where $k$ is an integer and $o(1)$ denotes a term that vanishes
asymptotically at large~$A$. The additive constant depends on the choice of
the inner product and, when the Hamiltonian is not essentially
self-adjoint, also on the choice of the self-adjoint extension.
With two particular choices for the inner product, we can verify (and
improve on) the accuracy of this WKB result rigorously; with another two
particular choices, we can rigorously verify the accuracy of the leading
order term.
We can therefore view our theory as producing, from a Hamiltonian quantum
theory constructed from first principles, the area spectrum
(\ref{area-quantization}) with $\alpha = 32\pi$.

We shall argue that the discreteness of the quantum spectrum is related to
the classical incompleteness of the theory. As the variable $a$
classically reaches the singularity at $a=0$ within finite parameter
time,  both in the past and in the future, the classical theory can be
thought of as particle motion on the positive half-line in a confining
potential. Whenever the self-adjoint Hamiltonian operator is constructed
so that the possible `quantum potential' part does not become significant,
general theorems guarantee that the spectrum of the Hamiltonian will be
discrete \cite{dunfordII,reed-simonII}. In physical terms, wave packets
following classical trajectories will be reflected quantum mechanically
from the origin, and the quantum dynamics will in this sense have a
quasiperiodic character. In contrast, if the spacetime foliation were
chosen so that it would take an infinite amount of parameter time for the
variable $a$ to reach the singularity, the classical theory could be
thought of as particle motion on the full real line in a potential that is
confining on the right but not on the left. In such potentials, the
spectrum of a self-adjoint Hamiltonian operator generically has a
continuous part, corresponding physically to the fact that wave packets
can travel arbitrarily far to the left without being reflected. We shall
present a simple example of each of these two types of foliation,
hand-picked so that the Hamiltonian operator becomes easily tractable.
{}From the first example we can reproduce the area spectrum
(\ref{area-quantization})  with an arbitrary value of the
constant~$\alpha$; in the second example, the spectrum of the Hamiltonian
operator will be continuous and consist of the full non-negative
half-line.

In addition to the above results for the vacuum theory, we shall
also briefly investigate the inclusion of a fixed electric charge and a
negative cosmological constant. With the analogous choices for the boundary
conditions, the phase space coordinates, and the Hamiltonian quantum
theory, we again show that the spectrum of the Hamiltonian operator is
discrete and bounded below. The distribution of the large eigenvalues
could presumably be analyzed by a suitable generalization of our
vacuum techniques; however, we shall not pursue this issue here.

The rest of the paper is as follows. In section
\ref{sec:throat-classical} we derive the reduced Hamiltonian theory in
our phase space variables, starting from Kucha\v{r}'s reduced phase space
variables \cite{kuchar1} and performing the appropriate
canonical transformation. The theory is quantized in
section~\ref{sec:throat-quantum}, with  considerable parts of the technical
analysis deferred to the three appendices. In section
\ref{sec:q-and-lambda} we discuss the inclusion of the electric charge
and a negative cosmological constant. Section \ref{sec:discussion}
contains a brief summary and a discussion.

For the remainder of this paper we shall work in natural units,
$\hbar = c = G = 1$.

\section{Classical wormhole throat theory}
\label{sec:throat-classical}

In this section we present a classical Hamiltonian theory of the
Schwarzschild black hole in terms of reduced phase space
variables that are associated with a wormhole throat, in a sense to be made
more precise below. We first briefly recall, in
subsection~\ref{subsec:kuchar-reduction}, Kucha\v{r}'s Hamiltonian
reduction of spherically symmetric vacuum geometries
\cite{kuchar1}. In subsection \ref{subsec:class-throat-derivation} we
derive the throat theory from Kucha\v{r}'s reduced theory via a suitable
canonical transformation.

\subsection{Kucha\v{r} reduction}
\label{subsec:kuchar-reduction}

We start from the general spherically symmetric ADM line element
\begin{equation}
ds^2 = - N^2 dt^2 + \Lambda^2 {(dr + N^r dt)}^2 +R^2 d\Omega^2
\ \ ,
\label{4-metric}
\end{equation}
where $d\Omega^2$ is the metric on the unit two-sphere, and $N$,
$N^r$, $\Lambda$, and $R$ are functions of $t$ and $r$ only.
We adopt the falloff conditions of Ref.\ \cite{kuchar1}. These conditions
render the spacetime asymptotically flat both at
$r\to\infty$ and $r\to-\infty$, and they make $|r|$ coincide asymptotically
with the spacelike radial proper distance coordinate in Minkowski space.
Each classical solution consists of some portion of the Kruskal
manifold \cite{MTW}, such that the  constant $t$ hypersurfaces extend from
the left hand side spacelike infinity to the right hand side spacelike
infinity, crossing the horizons in arbitrary ways. In particular, the
Schwarzschild mass is positive for every classical solution. The falloff
conditions also guarantee that the four-momentum at the infinities has no
spatial component: the black hole is at rest with respect to the left and
right asymptotic Minkowski frames.

We fix the asymptotic values of $N$ at $r\to\pm\infty$ to be prescribed
$t$-dependent quantities, denoted by~$N_\pm(t)$. The Hamiltonian form of the
Einstein action, with appropriate boundary terms, reads then
\begin{eqnarray}
S &=& \int dt
\int_{-\infty}^\infty dr \left( P_\Lambda {\dot \Lambda} +
P_R {\dot R}
- N{\cal H} - N^r {\cal H}_r
\right)
\nonumber
\\
&&- \int dt \left( N_+ M_+ + N_- M_- \right)
\ \ ,
\label{S-ham}
\end{eqnarray}
where ${\cal H}$ and ${\cal H}_r$ are respectively the super-Hamiltonian
constraint and the radial supermomentum constraint. The quantities
$M_\pm(t)$ are determined by the asymptotic falloff of the configuration
variables, and on a classical solution they both
are equal to the Schwarzschild
mass. We refer to Ref.\ \cite{kuchar1} for the details. Note that when
varying the action~(\ref{S-ham}), $N_\pm(t)$ are considered fixed.

Through a judiciously chosen canonical transformation, followed by
elimination of the constraints by Hamiltonian reduction, Kucha\v{r}
\cite{kuchar1} brings the action (\ref{S-ham}) to the unconstrained
Hamiltonian form
\begin{equation}
S = \int dt \left[ \bp {\dot \bm}
- \left( N_+ + N_- \right) \bm \right]
\ \ ,
\label{action1}
\end{equation}
where the independent variables $\bm$ and $\bp$ are functions of $t$ only.
The degrees of freedom constitute thus the single canonical pair
$(\bm,\bp)$: the theory is no longer that of fields, but that of finitely
many degrees of freedom. The variables take the values
$\bm>0$ and
$-\infty<\bp<\infty$, and their equations of motion are
\begin{mathletters}
\label{red-eom}
\begin{eqnarray}
{\dot \bm}
&=& 0
\ \ ,
\label{red-eom-m}
\\
{\dot \bp}
&=&
- N_+ - N_-
\ \ .
\label{red-eom-p}
\end{eqnarray}
\end{mathletters}%

For our purposes, it will not be necessary to recall the details of the
derivation of the action~(\ref{action1}), but what will be important is the
interpretation of the reduced theory in terms of the spacetime geometry. On
each classical solution, the time-independent value of $\bm$ is simply the
value of the Schwarzschild mass. By Birkhoff's theorem, $\bm$ thus carries
all the information about the local geometry of the classical solutions. The
variable~$\bp$, on the other hand, is equal to the difference of the
asymptotic Killing times between the left and right infinities on a
constant $t$ hypersurface, in the convention where the Killing time at the
right (left) infinity increases towards the future (past). The two terms in
the evolution equation (\ref{red-eom-p}) arise respectively from the two
infinities. Thus, $\bp$~contains no information about the local geometry,
but instead it carries the information about the anchoring of the spacelike
hypersurfaces at the two infinities. Note that~$\bp$, as the difference of
the asymptotic Killing times, is invariant under the global isometries that
correspond to translations in the Killing time.

\subsection{Hamiltonian throat theory}
\label{subsec:class-throat-derivation}

We shall now make two restrictions on the reduced
theory~(\ref{action1}). First, we specialize to
$N_+=1$ and $N_-=0$. This means that the parameter time $t$
coincides with the asymptotic right hand side Minkowski time, up to an
additive constant, whereas at the left hand side infinity the hypersurfaces
remain frozen at the same value of the asymptotic Minkowski time for
all~$t$. The action reads then
\begin{equation}
S = \int dt \left( \bp {\dot \bm}
- \bm \right)
\ \ .
\label{action2}
\end{equation}
The fact that the Hamiltonian now equals simply $\bm$ reproduces the
familiar identification of the Schwarzschild mass as the ADM energy, from
the viewpoint of asymptotic Minkowski time evolution at {\em one\/}
asymptotically flat infinity. Our choice as to which of the two infinities
has been taken to evolve is of course merely a convention, but the choice
of completely freezing the evolution at the other infinity arises from the
requirement that our theory describe physics accessible to observers at
just one infinity. We shall return to this issue in
section~\ref{sec:discussion}.

Second, for reasons that will become transparent below, we confine the
variables by hand to the range $|\bp|<\pi\bm$. This means that in each
classical solution, the asymptotic right hand side Minkowski time only
takes values within an interval of length~$2\pi\bm$, centered around a
value that is diagonally opposite to the non-evolving left end of the
hypersurfaces in the Kruskal diagram. In terms of the parameter time~$t$,
each classical solution is then  defined only for an interval
$-\pi\bm<t-t_0<\pi\bm$, where $t=t_0$ is the hypersurface whose two
asymptotic ends are diagonally opposite.

Consider now the transformation from the pair $(\bm,\bp)$ to the new
pair $(a,p_a)$ defined by the equations
\begin{mathletters}
\label{transformation}
\begin{eqnarray}
|\bp| &=&
\int_a^{2\bm} {db \over \sqrt{2\bm b^{-1} -1}}
\nonumber
\\
&=&
\sqrt{2\bm a - a^2}
+ \bm \arcsin\left( 1 - a / \bm \right) +  \casehalf \pi\bm
\ \ ,
\label{trans1}
\\
p_a &=&
{\rm sgn}(\bp) \, \sqrt{2\bm a - a^2}
\ \ .
\label{trans2}
\end{eqnarray}
\end{mathletters}%
The ranges of the variables are $a>0$ and $-\infty<p_a<\infty$.
The transformation is well-defined, one-to-one, and canonical. The new
action reads
\begin{equation}
S = \int dt \left( p_a {\dot a}
- H \right)
\ \ ,
\label{throat-action}
\end{equation}
where the Hamiltonian is given by
\begin{equation}
H = {1\over2} \left( {p_a^2 \over a} + a \right)
\ \ .
\label{throat-hamiltonian}
\end{equation}
The classical solutions are easily written out in the new variables. The
value of $H$ on a classical solution is just~$\bm$, and by writing the
canonical momentum $p_a$ in (\ref{throat-hamiltonian}) in terms of $a$
and~$\dot{a}$, one recovers the equation of motion for $a$ in the form
\begin{equation}
\dot{a}^2 = {{2\bm}\over a} - 1
\ \ .
\label{geodesic-eq}
\end{equation}
Hence the configuration variable $a$ starts from zero at $t = t_0 -
\pi\bm$, reaches the maximum value $2\bm$ at $t=t_0$, and collapses back
to zero at $t = t_0 + \pi\bm$.

The interest in the variables $(a,p_a)$ is that they have an appealing
geometrical interpretation in terms of the dynamics of a wormhole throat in
the black hole spacetime. To see this, we recall that the derivation
of the reduced action (\ref{action1}) from the original geometrodynamical
action (\ref{S-ham}) in Ref.\ \cite{kuchar1} relied on the properties of the
spacelike hypersurfaces only through their asymptotic behavior, but
otherwise left these hypersurfaces completely arbitrary. We can thus
exercise this freedom and seek an interpretation of the variables $(a,p_a)$
in terms of a suitably chosen foliation.

The crucial observation is now that equation (\ref{geodesic-eq}) is
identical to the equation of a radial timelike geodesic through the
bifurcation two-sphere of a Kruskal manifold of mass~$\bm$, {\em
provided\/} one identifies $a$ as the curvature radius of the two-sphere
and the dot as the proper time derivative. This is easily seen for example
from the expression of the interior Schwarzschild metric in the
Schwarzschild coordinates. With these identifications, equations
(\ref{transformation}) show  that $-\bp$ becomes identified with the proper
time elapsed along this geodesic from the bifurcation two-sphere, with
positive (negative) values of $-\bp$ yielding the part of the geodesic that
is in the black (white) hole interior. Thus, if there exists a foliation
consistent with the falloff conditions of Ref.\ \cite{kuchar1},
intersecting a timelike geodesic through the bifurcation two-sphere so
that $-\bp$ agrees with the proper time on the geodesic in this fashion,
then the quantity $a$ defined by (\ref{trans1}) is the two-sphere radius
along this geodesic.

It is easy to see that foliations of this kind do exist. Let us briefly
discuss an example that is closely related to the Novikov
coordinates \cite{MTW}. Recall that the geometric idea behind the Novikov
coordinates consists of fixing a spacelike hypersurface of constant
Killing time through the Kruskal manifold, and releasing from this
hypersurface a family of freely falling test particles with a vanishing
initial three-velocity in the Schwarzschild coordinates. The coordinates
$(\tau,R^*)$ are then defined so that they follow these test particles:
the trajectories are the lines of constant $R^*$, and on each trajectory
$\tau$ is equal to the proper time. The initial hypersurface is
$\tau=0$, with $R^*>0$ and $R^*<0$ giving respectively the halves living in
the right and left exterior regions. Now, to arrive at a foliation
satisfying our requirements, we first deform the Novikov coordinates near
the left hand side infinity to accommodate the condition $N_-=0$, and we
then redefine $R^*$ near the infinities in a $\tau$-independent fashion so
as to conform to the radial falloff assumed in Ref.\ \cite{kuchar1} (the
right hand side infinity will then have, in the notation of Ref.\
\cite{kuchar1}, a falloff with $\epsilon=1$). The distinguished geodesic
through the bifurcation two-sphere is given by $R^*=0$, and the coordinate
$\tau$ agrees by construction both with the proper time along this
geodesic and with the asymptotic Minkowski time at the right hand side
infinity.

Our interpretation of $a$ gives now a geometrical reason for the
restriction $|\bp|<\pi\bm$,  which we above introduced by hand. As a
radial timelike geodesic from the initial singularity to the final
singularity through the bifurcation two-sphere has the finite total proper
time~$2\pi\bm$, foliations satisfying our conditions do not cover all of
the spacetime. The foliations only exist for the duration of $2\pi\bm$ in
the asymptotic right hand side Minkowski time.\footnote{Note that although
the Novikov coordinates $(\tau,R^*)$ are globally well-defined on the
Kruskal manifold, the hypersurfaces of constant $\tau$ extend from one
infinity to the other only for $|\tau| < \pi\bm$. For larger values of
$|\tau|$, these hypersurfaces hit a singularity.}

We summarize. Fix a radial timelike geodesic through the bifurcation
two-sphere, and choose any foliation, consistent with our falloff
conditions, such that the proper time along the geodesic and the
asymptotic right hand side Minkowski time agree on the constant $t$
hypersurfaces. Then, the variable $a$ equals the radius of the two-sphere
on the distinguished geodesic. In particular, if the foliation is chosen
so that on each constant $t$ hypersurface, the radius of the two-sphere
attains its minimum value on the distinguished geodesic, then this
geodesic and the ones obtained from it by spherical symmetry form the
trajectory of the Einstein-Rosen wormhole throat. This is the case for
example in foliations obtained by deforming the Novikov coordinates near
the left hand side infinity in the manner discussed above. We shall
therefore, with a minor abuse of terminology, refer to $a$ as the radius
of the wormhole throat, and to the theory given by (\ref{throat-action})
and (\ref{throat-hamiltonian}) as the Hamiltonian throat theory.

It is important to note that the configuration variable $a$ is
bounded on each classical trajectory, reaching the maximum value $2\bm$ as
the wormhole throat crosses the bifurcation two-sphere. This means, in a
certain sense, that the spacetime dynamics in terms of our configuration
variable $a$ is confined `inside' the hole. This is physically appealing
from the viewpoint of an observer at infinity: such an observer sees the
exterior region of the spacetime as static.

\section{Throat quantization}
\label{sec:throat-quantum}

In this section we shall quantize the Hamiltonian throat theory of
section~\ref{sec:throat-classical}. We saw above that the
classical Hamiltonian is numerically equal to the
Schwarzschild mass, and that this Hamiltonian arises as the energy with
respect to the Minkowski time evolution at one asymptotically flat
infinity. The quantum Hamiltonian operator can therefore be viewed as the
energy operator with respect to an asymptotic Minkowski frame in which the
hole is at rest. In particular, the spectrum of the Hamiltonian operator
becomes the ADM mass spectrum of the hole. Our main aim will be a
qualitative analysis of this spectrum.

We take the states of the quantum theory to be described by functions
of the configuration variable~$a$. The Hilbert space is $\hilbert
:= L^2(\BbbR^+; \mu da)$, with the inner product
\begin{equation}
(\psi_1,\psi_2) = \int_0^\infty \mu da \, \overline{\psi_1} \psi_2
\ \ ,
\label{ip}
\end{equation}
where $\mu(a)$ is some smooth positive weight function. To obtain the
Hamiltonian operator~${\hat H}$, we make in the classical Hamiltonian
(\ref{throat-hamiltonian}) the substitution $p_a \to -i d/(da)$ and adopt a
symmetric ordering with respect to the inner product~(\ref{ip}). The result
is
\begin{equation}
{\hat H}
= {1\over2} \left[
- {1\over\mu}{d \over d a}
\left({\mu\over a} {d \over d a} \right) + a
\right]
\ \ .
\label{Hfirst}
\end{equation}

For technical reasons, it will be useful to work with an isomorphic
theory in which the inner product and the kinetic term of the
Hamiltonian take a more conventional form. To achieve this, we write
$a = x^{2/3}$, $\mu = \case{3}{2}x^{1/3}\nu^2$, and $\psi =
\nu^{-1}\chi$. The theory above
is then mapped to the theory whose Hilbert space is
$\hilbertzero
:= L^2(\BbbR^+; dx)$, with the inner product
\begin{equation}
{(\chi_1,\chi_2)}_0 = \int_0^\infty dx \, \overline{\chi_1} \chi_2
\ \ .
\label{ip0}
\end{equation}
The new Hamiltonian operator is
\begin{equation}
\hhzero
= {9\over8} \left[ - {d^2 \over dx^2} + {4 x^{2/3} \over9}
+ {\nu'' \over \nu}
\right]
\ \ ,
\label{hhzero-nu}
\end{equation}
where $'= d/dx$. By construction, $\hhzero$ is a symmetric operator
in~$\hilbertzero$.
Note that if we had retained dimensions, the `quantum potential' term
$\case{9}{8}\nu''/\nu$ would be proportional to~$\hbar^2$.

To completely specify the quantum theory, we need to make $\hhzero$ into
a self-adjoint operator on~$\hilbertzero$. The possible ways of doing this
depend on the quantum potential term $\case{9}{8}\nu''/\nu$. For
concreteness, we shall from now on take $\mu= a^s$, where $s$ is a real
parameter. The qualitative results would, however, be analogous for any
sufficiently similar $\mu$ with a power-law asymptotic behavior at $a\to0$
and $a\to\infty$.

With $\mu= a^s$, $\hhzero$ takes the form
\begin{equation}
\hhzero
= {9\over8} \left[ - {d^2 \over dx^2} + {4 x^{2/3} \over9}
+ {r(r-1)\over x^2}
\right]
\ \ ,
\label{hhzero}
\end{equation}
where
\begin{mathletters}
\label{r-def}
\begin{eqnarray}
r &=&
(2s-1)/6
\qquad \hbox{for $s\ge2$}
\ \ ,
\label{r-def-1}
\\
r &=& (7-2s)/6
\qquad \hbox{for $s<2$}
\ \ .
\end{eqnarray}
\end{mathletters}%
We could have replaced (\ref{r-def}) by (say) (\ref{r-def-1}) for all~$s$,
with (\ref{hhzero}) still holding. However, as (\ref{hhzero}) is
invariant under $r\to 1-r$, it will be sufficient to analyze $\hhzero$
for $r\ge 1/2$; this range for $r$ is recovered through the
definition~(\ref{r-def}).

It is easy to see that $\hhzero$ has self-adjoint
extensions for any~$r$. In the terminology of Ref.\ \cite{reed-simonII},
infinity is a limit point case, whereas zero is a limit point case for
$r\ge 3/2$ and limit circle case otherwise (Ref.\ \cite{reed-simonII},
Theorems X.8 and X.10, and Problem~7). For $r\ge 3/2$, $\hhzero$ is
therefore essentially self-adjoint. For $1/2 \le r < 3/2$, on the other
hand, the self-adjoint extensions of $\hhzero$ are characterized by a
boundary condition at zero and parametrized by~$U(1)$.

We now wish to extract qualitative information about the
spectrum of the self-adjoint extensions of~$\hhzero$.

A first observation is that the essential spectrum of every
self-adjoint extension of $\hhzero$ is empty (Ref.\ \cite{dunfordII},
Theorems XIII.7.4, XIII.7.16, and XIII.7.17). This means that the
spectrum is discrete:
the spectrum consists of eigenvalues corresponding to  genuine,
normalizable eigenstates, and the eigenvalues have disjoint neighborhoods.

Second, we shall show in appendix \ref{app:semi-boundedness} that every
self-adjoint extension of $\hhzero$ is bounded below: the system has a
ground state. For $r\ge 3/2$, the ground state energy is always positive.
For $1/2 \le r < 3/2$, the ground state energy depends on the
self-adjoint extension, and the situation is more versatile. On the one
hand, there is a certain (open) set among the self-adjoint extensions within
which the ground state energy is positive. On the other hand, there exist
extensions whose ground state energy is arbitrarily negative.

Third, we shall show in appendix \ref{app:large-eigen} that a WKB
analysis yields for the squares of the large eigenenergies the
asymptotic estimate
\begin{equation}
E_{\rm WKB}^2 \sim 2k + \hbox{constant} + o(1)
\ \ ,
\label{asymptotic-energy}
\end{equation}
where $k$ is an integer and $o(1)$ denotes a term that vanishes
asymptotically at large~$E$. The constant depends on $r$ and, for
$1/2<r<3/2$, also on the self-adjoint extension, in a way discussed in
appendix~\ref{app:large-eigen}. We shall also be able to rigorously verify
the accuracy of this WKB result in the special case $r=5/6$, and the
accuracy of its leading order term in the special case $r=1$.

These properties of the spectrum have consequences of direct physical
interest. At the low end of the spectrum, the fact that the Hamiltonian is
bounded below indicates stability: one cannot extract from the system an
infinite amount of energy. At the high end of the spectrum, the asymptotic
distribution of the large eigenenergies yields for the black hole area the
eigenvalues~(\ref{area-asymptotic}): this agrees with
Bekenstein's area spectrum~(\ref{area-quantization}),
with $\alpha=32\pi$. We shall discuss the physical implications of
these results further in section~\ref{sec:discussion}.

\section{Throat theory with charge and a negative cosmological constant}
\label{sec:q-and-lambda}

In this section we shall outline how the throat theory can be generalized
to accommodate  electric charge and a negative cosmological constant. The
classical black hole solutions are in this case given (locally)
by the \rnads\ metric \cite{exact-book}
\begin{mathletters}
\label{rnads-solution}
\begin{equation}
ds^2 = - \left(
1 - {2M \over R} + {Q^2 \over R^2} + {R^2 \over \ell^2}
\right)
dT^2
+
{dR^2 \over
{\displaystyle{
1 - {2M \over R} + {Q^2 \over R^2} + {R^2 \over \ell^2}
}}
}
+ R^2 d\Omega^2
\ \ ,
\label{rnads-metric}
\end{equation}
with the electromagnetic potential
one-form
\begin{equation}
\bbox{A} = {Q \over R} \bbox{d}T
\ \ .
\label{rnads-A}
\end{equation}
\end{mathletters}%
The parameters $M$ and $Q$ are referred to as the mass and the (electric)
charge, and the cosmological constant has been written in terms of the
positive parameter $\ell$ as~$-3\ell^{-2}$. For the global structure of
the spacetime, see Refs.\ \cite{lake1,btz-cont}. We shall understand the
case of a vanishing cosmological constant as the limit $\ell\to\infty$,
and in this case the above solution reduces to the Reissner-Nordstr\"om
solution.

The Hamiltonian structure of the spherically symmetric Einstein-Maxwell
system with a cosmological constant was analyzed by a technique related to
Ashtekar's variables in Refs.\ \cite{thiemann3,thiemann4}. An analysis via
a Kucha\v{r}-type canonical transformation and Hamiltonian reduction,
both with and without a negative cosmological constant, was given in
Ref.\ \cite{lou-win}. Although the focus of Ref.\ \cite{lou-win} was on
thermodynamically motivated boundary conditions, which confine the constant
$t$ hypersurfaces to one exterior region of the spacetime, the
discussion therein generalizes without essential difficulty to boundary
conditions that allow the constant $t$ hypersurfaces to extend
from a left hand side spacelike infinity to the corresponding right hand
side spacelike infinity, crossing the event horizons in arbitrary ways but
crossing no inner horizons. The new technical issues arise mainly from
the fact that with a negative cosmological constant, the left and right
infinities are asymptotically \ads\ rather than asymptotically flat. The
new physical issues arise mainly in the choice of the electromagnetic
boundary conditions at the infinities.

We shall here concentrate on the theory where the electric charge is fixed
at the infinities. In the reduced theory, the charge then becomes an
entirely nondynamical, external parameter, which we denote by~$\bq$. On a
classical solution, $\bq$~is equal to $Q$
in~(\ref{rnads-solution}).

The action of the reduced theory takes the form~(\ref{action1}). On the
classical solutions, $\bm$ is equal to the mass parameter $M$
of~(\ref{rnads-metric}). The canonical conjugate $\bp$ again
equals the difference in the asymptotic Killing times between the left and
right ends of a constant $t$ hypersurface. The range of $\bm$ is
$\bm>\bm_{\rm crit}$, where the critical value $\bm_{\rm crit}(\bq,\ell)$
is positive for $\bq\ne0$ and vanishes for
$\bq=0$: this restriction arises from the requirement that the
classical solutions have a nondegenerate event horizon
\cite{lake1,btz-cont,lou-win}. The range of $\bp$ is $-\infty < \bp <
\infty$. The quantities $N_\pm$ determine the evolution of the ends of
the hypersurface in the asymptotic Killing time. For $\ell\to\infty$,
$N_\pm$ are simply the asymptotic values of the lapse, whereas for
$0<\ell<\infty$ they are related to the lapse by a factor that diverges at
the infinities. Note that in the special case of $\bq=0$ and
$0<\ell<\infty$, we get a theory of vacuum spacetimes with a negative
cosmological constant.

Mimicking section~\ref{sec:throat-classical}, we freeze the
evolution of the hypersurfaces at the left hand side infinity by setting
$N_-=0$, and fix the evolution at the right hand side infinity to proceed
at unit rate with respect to
the Killing time by setting $N_+=1$. The action is
given by~(\ref{action2}). The value of the Hamiltonian on a classical
solution is then equal to the mass parameter. For $\ell\to\infty$ this
reproduces the identification of the mass as the ADM energy from the
viewpoint of asymptotic Minkowski time evolution at one infinity, just as
in the uncharged case in section~\ref{sec:throat-classical}. For
$0<\ell<\infty$, we similarly recover the interpretation of the mass
parameter as the ADM-type energy from the viewpoint of asymptotic \ads\
Killing time evolution at one infinity.

In analogy with~(\ref{transformation}), we
introduce the new variables $(a,p_a)$ via the transformation
\begin{mathletters}
\label{lq-transformation}
\begin{eqnarray}
|\bp| &=&
\int^{a_+}_a
{db \over \sqrt{ 2 \bm b^{-1} - 1 - \bq^2 b^{-2} - b^2 \ell^{-2}
}}
\ \ ,
\label{lq-trans1}
\\
p_a &=&
{\rm sgn}(\bp) \, \sqrt{2\bm a - a^2 - \bq^2 - a^4 \ell^{-2}}
\ \ ,
\label{lq-trans2}
\end{eqnarray}
\end{mathletters}%
where $a_+(\bm,\bq,\ell)$ is the unique positive zero of the right hand
side in (\ref{lq-trans2}) for $\bq=0$, and the larger of the two positive
zeroes for $\bq\ne0$. On a classical solution, $a_+$~is the radius of the
event horizon. To make this transformation  well-defined, we again need to
restrict by hand the range of~$\bp$. For $\bq=0$, the upper limit for
$|\bp|$ is obtained from (\ref{lq-trans1}) with $a=0$. For
$\bq\ne0$, the upper limit for $|\bp|$ is obtained from
(\ref{lq-trans1}) with $a=a_-$, where $a_-(\bm,\bq,\ell)$ is the smaller of
the two positive zeroes of the right hand side in~(\ref{lq-trans2}). On a
classical solution, $a_-$ is the radius of the inner horizon.

With this restriction on the range of~$\bp$, the transformation
(\ref{lq-transformation}) is well-defined, one-to-one, and canonical. The
new action is given by (\ref{throat-action}) with the Hamiltonian
\begin{equation}
H = {1\over2} \left( {p_a^2 \over a} + a + {\bq^2 \over a} +
{a^3 \over \ell^2}\right)
\ \ ,
\label{lq-throat-hamiltonian}
\end{equation}
and the value of this Hamiltonian on a classical solution is just the
mass.

As in section~\ref{sec:throat-classical}, the theory has an
interpretation in terms of a wormhole throat. If there exists a foliation
with the appropriate falloff conditions \cite{lou-win},  intersecting a
timelike geodesic through the event horizon bifurcation two-sphere so
that $-\bp$ coincides with the proper time from the bifurcation two-sphere
along this geodesic, then the quantity $a$ defined by (\ref{lq-trans1}) is
the two-sphere radius on this geodesic. If the  foliation is chosen
suitably symmetric near the specified geodesic, we can think of $a$ as
the radius of the wormhole throat. We shall now examine the existence of
such foliations for the different values of the parameters $\bq$
and~$\ell$.

For $\bq=0$ and $0<\ell<\infty$, the classical solution is the
Schwarzschild-anti-de~Sitter hole. The Penrose diagram differs from that
of the Kruskal manifold only in that the asymptotically flat
infinities are replaced by asymptotically \ads\ infinities,
represented by vertical lines \cite{lake1,btz-cont}. Foliations of the
desired kind clearly exist: the timelike geodesic starts at the initial
singularity with $a=0$, reaches the maximum value of $a$ at the bifurcation
two-sphere, and ends at the final singularity with $a=0$. The situation is
thus qualitatively very similar to that with the Schwarzschild hole.

For $\bq\ne0$ and $\ell\to\infty$, the classical solution is the
Reissner-Nordstr\"om hole with $\bm>|\bq|$. The Penrose diagram can be found
in Refs.\ \cite{MTW,haw-ell}. Our Kucha\v{r}-type Hamiltonian formulation
is valid for the part of the spacetime that consists of one pair of
spacelike-separated left and right asymptotically flat regions and the
connecting region that is bounded in the past and future by the Cauchy
horizons. The solutions to the equations of motion obtained from the
Hamiltonian (\ref{lq-throat-hamiltonian}) are periodic oscillations in the
interval
$a_-\le a \le a_+$, but our derivation of this Hamiltonian is only valid
on each solution between two successive minima of~$a$. Now, it is clear that
foliations of the desired kind exist: the timelike  geodesic starts with
$a=a_-$ at the past Cauchy horizon bifurcation two-sphere, reaches $a=a_+$
at the event horizon bifurcation two-sphere, and ends with $a=a_-$ at the
future Cauchy horizon bifurcation two-sphere.

Finally, for $\bq\ne0$ and $0<\ell<\infty$, the classical solution is the
\rnads\ hole, with $\bm$ so large that a nondegenerate event horizon
exists. The Penrose diagram is obtained from that of the
Reissner-Nordstr\"om hole by replacing the asymptotically flat infinities
by asymptotically \ads\ infinities \cite{lake1,btz-cont}. Our
Kucha\v{r}-type Hamiltonian formulation is valid for the part of the
spacetime that consists of one pair of spacelike-separated  left and right
asymptotically \ads\ regions and the connecting region that is bounded in
the past and future by the inner horizons. The inner horizons are now not
Cauchy horizons, as the asymptotically \ads\ infinities render our part
of  the spacetime not globally hyperbolic.  As with the
Reissner-Nordstr\"om hole above, the solutions to the equations of motion
obtained from the Hamiltonian (\ref{lq-throat-hamiltonian}) are periodic
oscillations in the interval $a_-\le a \le a_+$, and our derivation of
this Hamiltonian is only valid on each solution between two successive
minima of~$a$. Foliations of the desired kind now exist while the timelike
geodesic remains sufficiently close to the event horizon bifurcation
two-sphere. However, it is seen from the Penrose diagram that as the
geodesic progresses towards the past and future inner horizon bifurcation
two-spheres, there will occur a critical value of the proper time after
which the constant $t$ hypersurfaces would necessarily need to become
somewhere timelike. Therefore, the throat interpretation can only be
maintained in the full domain of validity of the Hamiltonian
(\ref{lq-throat-hamiltonian}) by appealing to a foliation where the
constant $t$ hypersurfaces  need not be everywhere spacelike. We shall
return to this issue in section~\ref{sec:discussion}.

Quantization of the theory proceeds as in
section~\ref{sec:throat-quantum}. The Hamiltonian $\hhzero$
(\ref{hhzero}) inherits the additional terms
\begin{equation}
{\bq^2 \over 2 x^{2/3}}
+
{x^2 \over 2 \ell^2}
\ \ .
\end{equation}
The theorems cited in section \ref{sec:throat-quantum} show that the
additional terms make no difference for the existence and counting of the
self-adjoint extensions of~$\hhzero$, and they also show that the
essential spectrum of any self-adjoint extension of $\hhzero$ is again
empty. For $\bq=0$, the proof of the lower bound for the spectrum given in
appendix \ref{app:semi-boundedness} goes through virtually without change.
For $\bq\ne0$, the charge term modifies the small $x$ behavior of the wave
functions, and the analysis of the self-adjointness boundary condition is
more involved; in particular, there is a qualitative difference between
the cases $7/6 < r < 3/2$, $r = 7/6$, and $1/2 < r < 7/6$, arising from
whether the next-to-leading term in the counterpart of $v_E(x)$ in
(\ref{uv}) dominates the leading order term in the counterpart of $u_E(x)$
at small~$x$. However, the modified Bessel function asymptotic behavior
(\ref{chi-as-negE}) and (\ref{chi-as-negE-crit}) still holds, as can be
shown by applying the series solution method for $u_E$ and the `second
solution' integral formula for $v_E$ (see, for example, Chapter 8 of Ref.\
\cite{arfken}). The spectrum of every self-adjoint extension is therefore
again bounded below, and certain self-adjoint extensions are strictly
positive.

One expects that the asymptotic distribution of the large eigenvalues
could be investigated via a suitable generalization of the WKB techniques of
appendix~\ref{app:large-eigen}. We shall, however, not attempt to carry
out such an analysis here.

\section{Discussion}
\label{sec:discussion}

In this paper we have considered a Hamiltonian theory of spherically
symmetric vacuum Einstein gravity under Kruskal-like boundary conditions.
The foliation was chosen such that the evolution of the spacelike
hypersurfaces is frozen at the left hand side infinity, but proceeds at
unit rate with respect to the asymptotic Minkowski time at the right hand
side infinity.  The reduced Hamiltonian theory was written in a set of
variables associated with the Einstein-Rosen wormhole throat: the
configuration variable is the radius of the throat, in a foliation in
which the proper time at the throat agrees with the asymptotic right hand
side Minkowski time. The classical Hamiltonian is numerically equal to the
Schwarzschild mass.

We quantized the theory by Hamiltonian methods, taking the wave functions
to be functions of the classical configuration variable, and including a
general power-law weight factor in the inner product. The classical
Hamiltonian was promoted into a self-adjoint Hamiltonian operator. We
found that the spectrum of the Hamiltonian operator is discrete and
bounded below for all the
choices of the weight factor. In the cases where the
Hamiltonian operator is essentially self-adjoint, the spectrum is
necessarily positive definite; in the remaining cases, self-adjoint
extensions with a positive definite spectrum always exist. In all the
cases, a WKB estimate gave for the large eigenvalues the asymptotic
behavior~$\sqrt{2k}$, where $k$ is an integer, and we were able to
rigorously verify the accuracy of this estimate for four
particular choices of the weight factor. The resulting spectrum for the area
of the black hole agrees with the spectrum (\ref{area-quantization})
proposed by Bekenstein and others, with the dimensionless
constant $\alpha$ taking the value~$32\pi$. We also showed that analogous
results can be obtained in the presence of a fixed electric charge and a
negative cosmological constant.

It is perhaps worth emphasizing that the basic postulates of our quantum
theory consisted of the choice of a Hilbert space and a self-adjoint
Hamiltonian operator on it. We did not attempt to define more
`elementary' operators, such as those for `position' or `momentum',
self-adjoint or otherwise, from which the Hamiltonian operator could be
constructed. This issue might merit further study within some geometric or
algebraic framework of  quantization
\cite{woodhouse,isham-LH,AAbook}.\footnote{For example, note that the
kinetic term of the Hamiltonian operator
${\hat H}$ (\ref{Hfirst}) can be written as $\case{1}{2} \mu^{-1/2} {\hat
p}_a a^{-1}\mu {\hat p}_a \mu^{-1/2}$, where ${\hat p}_a: \psi \mapsto -i
\mu^{-1/2} (d/da) \mu^{1/2}\psi$ is a symmetric (but not self-adjoint)
momentum operator that can be regarded as conjugate to the position
operator ${\hat a}: \psi \mapsto a \psi$. We thank Thomas Strobl for this
observation.}

Even though our theory is that of pure vacuum, our boundary conditions
were chosen so as to make the results relevant for physics that is
accessible to an inertial observer at a spacelike infinity. Our spacelike
hypersurfaces have evolution at only {\em one\/} infinity,
and there they evolve at unit rate with respect to the asymptotic
Minkowski time. Our classical Hamiltonian is therefore the gravitational
Hamiltonian with respect to the proper time of an inertial observer at the
infinity, at rest with respect to the hole. It is thus reasonable
to think of the eigenvalues of the Hamiltonian operator as the possible
outcomes that an asymptotic observer would in principle obtain when
measuring the ADM mass of the hole. In a given (pure) quantum state, the
probability for obtaining a given eigenvalue is determined by the component
of the state in the respective eigenspace in the standard way. Although we
are here for concreteness using language adapted to a Copenhagen-type
interpretation, a translation into interpretations of the many-worlds
type  could easily be made.

One can also make a case that our throat variable $a$ depicts in a
particularly natural way the dynamical aspects of the Kruskal
manifold. Classically, the wormhole throat begins life at the white hole
singularity, expands to maximum radius at the bifurcation two-sphere, and
collapses to the black hole singularity. The dynamics of $a$ is
therefore, in a certain sense, confined to the interior regions of the
Kruskal manifold, and these are precisely the regions that do not admit
a timelike Killing vector. {}From a physical viewpoint, using a variable
with this property is motivated by the fact that an inertial observer at a
spacelike infinity sees her exterior region of the Kruskal manifold as
static.  Further, our foliation made the proper time at the throat increase
at the same rate as the asymptotic right hand side Minkowski time: by the
principle of equivalence, one may see this as the preferred condition for
relating the experiences of an inertial observer at the asymptotic
infinity to the experiences of an inertial observer at the throat. We
recall, in contrast, that the reduced phase space variables of Refs.\
\cite{thiemann1,thiemann2,kuchar1} reflect more closely  the {\em
static\/} aspects of the Kruskal manifold. Yet another set of variables
has been discussed in Refs.\ \cite{cava1,cava2}.

With this physical picture, the properties we obtained for the spectrum of
the Hamiltonian operator acquire consequences of direct physical interest.
At the low end of the spectrum, the fact that the Hamiltonian is bounded
below indicates stability: one cannot extract from the system an infinite
amount of energy. At the high end of the spectrum, in the semiclassical
regime of the theory, the discreteness of the spectrum in accordance with
the area quantization rule (\ref{area-quantization}) yields the
macroscopically observable consequences discussed by Bekenstein and
Mukhanov \cite{bek-mu}. It should be emphasized, however, that these
arguments operate at a somewhat formal level, as our theory does not
describe how the quantum black hole would interact with other degrees of
freedom, such as departures from spherical symmetry or matter fields.

On the grounds of classical positive energy theorems, one may feel inclined
to exclude by fiat quantum theories in which the ground state energy of an
isolated self-gravitating system is negative. Among our theories, this
would amount to a restriction on the self-adjoint extension in the cases
where the Hamiltonian operator is not essentially self-adjoint. However,
given the freedom that we have already allowed in the choice of the the
inner product, it would be a relatively minor further generalization to add
to our Hamiltonian operator the identity operator with some real
coefficient, and to take the coefficient as a new parameter in the quantum
theory.\footnote{We thank John Friedman for this observation.} The
classical limit of the quantum theory would still be correct, provided the
new parameter is understood to be proportional to Planck's constant. When
the Hamiltonian operator is not essentially self-adjoint, any given
self-adjoint extension can then be made positive definite by choosing the
new parameter sufficiently large. Note, however, that with any fixed value
of the new parameter, there still exist self-adjoint extensions whose
ground state energy is arbitrarily negative.

In section~\ref{sec:throat-classical}, we obtained the classical
Hamiltonian throat theory by first going from the geometrodynamical
Hamiltonian variables to Kucha\v{r}'s reduced Hamiltonian theory, and
then performing a suitable canonical transformation. The interpretation of
our variable $a$ as the radius of the wormhole throat was only introduced
after the fact, by appealing to a particular choice of the spacetime
foliation. We took the same route in the presence of charge and a negative
cosmological constant in section~\ref{sec:q-and-lambda}. We have not
discussed here how to derive a throat theory directly from the unreduced
geometrodynamical Hamiltonian theory by introducing a gauge and performing
the Hamiltonian reduction, but with the appropriate gauge choice,
the resulting theory should by construction be at least locally identical to
ours. The only case where one anticipates a difference in the global
properties is in the presence of both a nonvanishing charge and a negative
cosmological constant. In this case, we saw in section
\ref{sec:q-and-lambda} that while our throat theory is valid on each
classical solution between two successive minima of~$a$, the wormhole
throat interpretation could be maintained for all of this interval only by
appealing to a foliation that is not everywhere spacelike. A~direct
Hamiltonian reduction of the geometrodynamical theory in the corresponding
wormhole-type gauge would thus only yield our theory in a more limited
domain, valid on each classical solution in a certain interval around a
maximum of~$a$.

Our choice of freezing the evolution of the spacelike hypersurfaces at
the left hand side infinity was motivated by the desire to have a theory
that would describe physics accessible to observers at just one infinity.
For a vanishing charge and cosmological constant, this motivation can be
implemented at the very beginning by setting up the Hamiltonian theory not
on the Kruskal manifold, but instead on the
$\RPthree$ geon \cite{topocen}. To see this, recall \cite{topocen} that
the $\RPthree$ geon is the quotient of the Kruskal manifold under a freely
acting involutive isometry: this isometry consists of a reflection of the
Kruskal diagram about the vertical timelike line through the bifurcation
point, followed by the antipodal map on the two-sphere. The $\RPthree$
geon has thus only one exterior region, identical to one of the
Kruskal exterior regions. Further, the $\RPthree$ geon possesses a
distinguished $\RPtwo$ of timelike geodesics through the image of the
Kruskal bifurcation two-sphere: in the Penrose diagram \cite{topocen},
these geodesics go straight up along the `boundary' of the diagram. The
existence of the distinguished geodesics reflects the fact that
translations in the Killing time on the Kruskal manifold do not descend
into globally defined isometries of the $\RPthree$ geon. Now, Kucha\v{r}'s
canonical transformation and Hamiltonian reduction generalize readily to
the $\RPthree$ geon
\cite{LW2}. The reduced action is obtained from (\ref{action1}) by setting
$N_-=0$, and the momentum $\bp$ is now equal to the difference of the
Killing times between the distinguished timelike geodesics and the
single spacelike infinity. Setting $N_+=1$, we are led to the
action (\ref{action2}). The variable $a$ defined by (\ref{transformation})
is now equal to the curvature radius of the distinguished
$\RPtwo$ of timelike geodesics.

The above interpretation of our theory in terms of the $\RPthree$ geon
generalizes immediately to accommodate a negative cosmological constant.
For a nonvanishing charge, on the other hand, there exists again an
analogous involutive isometry that can be used to quotient the manifold,
but the electric field is invariant under this isometry only up to its
sign. This reflects the fact that Gauss's theorem prohibits a regular
spacelike hypersurface with just one asymptotic infinity from carrying a
nonzero charge. The $\RPthree$ geon interpretation does therefore not
extend to the charged case with a conventional implementation of the
electromagnetic field.

As we have seen, the central input in our classical theory was to
parametrize the geometry in terms of the radius of the wormhole throat in
a judiciously chosen foliation: our variable $a$ is the two-sphere radius
on a radial geodesic through the event horizon bifurcation two-sphere, in a
foliation such that the proper time along the distinguished geodesic
agrees with the asymptotic Killing time at the right hand side infinity.
One possible generalization would be to relax the requirement that the
variable `live' in the interior regions of the manifold, and use
instead timelike geodesics that do not pass through the bifurcation
two-sphere. To examine this, let us for concreteness set the charge and the
cosmological constant to zero, and let us generalize the canonical
transformation (\ref{transformation}) to
\begin{mathletters}
\label{transformation-gen}
\begin{eqnarray}
|\bp| &=&
\int\limits_a^{2\bm{\left[1 + {(R^*_0)}^2 \right]}}
{db \over
\sqrt{2\bm b^{-1}
-
{\left[1 + {(R^*_0)}^2 \right]}^{-1}
}}
\ \ ,
\\
\noalign{\medskip}
p_a &=&
{\rm sgn}(\bp) \,
\sqrt{2\bm a
-
a^2 {\left[1 + {(R^*_0)}^2 \right]}^{-1}
}
\ \ ,
\end{eqnarray}
\end{mathletters}%
where $R^*_0$ is a real-valued parameter. The transformation
(\ref{transformation}) is recovered as the special case $R^*_0=0$. The
ranges of the new variables are again $a>0$ and $-\infty<p_a<\infty$, but
the restriction for $\bp$ is now
$|\bp| < \pi  {\left[1 + {(R^*_0)}^2 \right]}^{3/2} \bm$. The action is
given by (\ref{throat-action}) with the Hamiltonian
\begin{equation}
H = {1\over2} \left[
{
{p_a^2 \over a}
+
{a
\over
1 + {(R^*_0)}^2_{\vphantom{A}}
}
}_{\vphantom{A_a}}
\right]
\ \ .
\label{throat-hamiltonian-gen}
\end{equation}
On a classical solution, the variable $a$ is now equal to the two-sphere
radius on a radial timelike geodesic whose trajectory is given by $R^* =
R^*_0$, where $R^*$ is the Novikov space coordinate in the notation of
Ref.\ \cite{MTW}. Foliations that make this
interpretation possible clearly exist. Simple examples are obtained by
deforming the Novikov foliation near the left hand side infinity as in
section~\ref{sec:throat-classical}, to accommodate the boundary condition
$N_-=0$, and (for $R^*_0\ne0$) also near the throat, to prohibit the
constant $t$ hypersurfaces from reaching the singularity before the
geodesic at $R^* = R^*_0$. Upon quantization along the lines of
section~\ref{sec:throat-classical}, we find that the spectrum depends on
the parameter $R^*_0$ only through an overall factor: if the eigenvalues
are denoted by $E_k^{(R^*_0)}$, where $k$ ranges over the nonnegative
integers, we have
\begin{equation}
E_k^{(R^*_0)}
=
{\left[1 + {(R^*_0)}^2 \right]}^{-3/4}
E_k^{(0)}
\ \ .
\end{equation}

A much wider generalization would be to relax the requirement, which we
above motivated by the equivalence principle, that the proper time along
the throat trajectory agree with the asymptotic Killing time.
If one allows this freedom, it is not difficult to come up with examples of
foliations in which the Hamiltonian takes a mathematically simple form. As
an illustration, let us exhibit two examples in the case of vanishing
charge and cosmological constant.

As the first example, suppose that $\bp$ is restricted by hand to have
the range $|\bp|<\gamma^{-1}\pi\bm$, where $\gamma$ is a positive constant.
We perform the canonical transformation
\begin{mathletters}
\begin{eqnarray}
\xi &=& \sqrt{2/\gamma} \, \bm
\cos \! \left( \gamma \bp \over 2\bm \right)
\ \ ,
\label{xi-def}
\\
\noalign{\smallskip}
p_\xi &=& \sqrt{2/\gamma} \,  \bm
\sin \! \left( \gamma \bp \over 2\bm \right)
\ \ ,
\end{eqnarray}
\end{mathletters}%
where the ranges of the new canonical variables are $\xi>0$ and
$-\infty<p_\xi <
\infty$.
The Hamiltonian takes the form
\begin{equation}
H = \sqrt{\casehalf \gamma ( p_\xi^2 + \xi^2 ) }
\ \ .
\end{equation}
We can identify $\sqrt{2\gamma}\,\xi$ as the two-sphere radius on a
radial geodesic through the bifurcation two-sphere in a foliation where the
proper time $\tau$ along this geodesic is
\begin{equation}
\tau =
- {\rm sgn}(\bp)
\int\limits_{\sqrt{2\gamma}\,\xi
}^{2\bm}
{db \over
\sqrt{2\bm b^{-1}
- 1
}}
\ \ ,
\end{equation}
with $\xi$ given by~(\ref{xi-def}).

To quantize this theory, we adopt the inner product
$(\psi_1,\psi_2) = \int_0^\infty d\xi \, \overline{\psi_1} \psi_2$. We
define the Hamiltonian operator ${\hat{H}}$ by spectral analysis as the
positive square root of some positive definite self-adjoint extension of
$\gamma {\hat{H}}_{\rm SHO}$, where ${\hat{H}}_{\rm SHO} := \casehalf
\left[ -{(d/d\xi)}^2 + \xi^2\right]$ is the simple harmonic oscillator
Hamiltonian operator on the positive half-line. The following
statements about ${\hat{H}}_{\rm SHO}$ can now be verified:
(i)
the self-adjoint extensions are specified by the boundary condition
$\cos(\theta) \psi - \sin(\theta) d\psi/d\xi =0$ at the origin, with
the parameter $\theta$ satisfying $0\le
\theta < \pi$;
(ii)
the spectrum of each self-adjoint extension
is purely discrete;
(iii)
the eigenfunctions are parabolic cylinder functions
\cite{Ab-Steg}, and for $\theta=0$ and $\theta=\pi/2$ they reduce
respectively to the odd and even ordinary harmonic oscillator wave
functions;
(iv)
if $\epsilon_k$ denotes the eigenvalues, with $k$ ranging over the
nonnegative integers, we have for $\theta=0$ and $\theta=\pi/2$ the
respective exact results
$\epsilon_k = 2k + \case{3}{2}$ and
$\epsilon_k = 2k + \casehalf$,
and for other values of $\theta$ the
asymptotic large $k$ expansion $\epsilon_k \sim 2k + \casehalf + \pi^{-1}
\cot(\theta) k^{-1/2} + o \! \left(k^{-1/2}\right)$;
(v)
the absence of negative eigenvalues is equivalent to the condition that
$\theta$ not lie in the interval $- 2^{-3/2}\pi^{-1}
{\left[ \Gamma(1/4) \right]}^2 < \tan(\theta) < 0$. The resulting spectrum
for ${\hat{H}}$ therefore agrees asymptotically with
the area quantization rule (\ref{area-quantization}) for any $\theta$ that
makes ${\hat{H}}_{\rm SHO}$ positive definite. The numerical constant
$\alpha$ takes the value~$32\pi\gamma$.

As the second example, suppose that $\bp$ retains the full range
$-\infty < \bp< \infty$, and perform the canonical transformation
\begin{mathletters}
\begin{eqnarray}
\eta &=&
{\bm^2 \over
\cosh \! \left(
{\displaystyle{
{\bp \over 2\bm}
}}
\right)
}
\ \ ,
\label{eta-def}
\\
\noalign{\medskip}
p_\eta &=&
\sinh \! \left(
{\displaystyle{
{\bp \over 2\bm}
}}
\right)
\ \ ,
\end{eqnarray}
\end{mathletters}%
where the ranges of the new canonical variables are $\eta>0$ and
$-\infty<p_\eta <
\infty$.
The Hamiltonian takes the form
\begin{equation}
H =
{(\eta^2 p_\eta^2 + \eta^2 )}^{1/4}
\ \ .
\end{equation}
We can identify $2\sqrt{\eta}$ as the two-sphere radius on a
radial geodesic through the bifurcation two-sphere in a foliation where the
proper time $\tau$ along this geodesic is
\begin{equation}
\tau =
- {\rm sgn}(\bp)
\int\limits_{2\sqrt{\eta}}^{2\bm}
{db \over
\sqrt{2\bm b^{-1}
- 1
}}
\ \ ,
\end{equation}
with $\eta$ given by~(\ref{eta-def}).

In the quantum theory we now adopt the inner product
$(\psi_1,\psi_2) = \int_0^\infty \eta^{-1} d\eta \, \overline{\psi_1}
\psi_2$. The operator ${\hat{H}}_{\rm exp} :=
- {[ \eta (d/d\eta) ]}^2 + \eta^2 $ is essentially self-adjoint, its
discrete spectrum is
empty, and its essential spectrum consists of the non-negative
half-line.\footnote{These statements follow in a straightforward
manner from Chapter VIII of  Ref.\ \cite{dunfordII}
after bringing ${\hat{H}}_{\rm exp}$ and the inner product to a standard
form by the substitution $\eta = e^\zeta$. Note that if we had retained
dimensions, this substitution would need to include a dimensionful
constant.}
We can therefore define the Hamiltonian operator ${\hat{H}}$ by spectral
analysis as ${({\hat{H}}_{\rm exp})}^{1/4}$.
It follows that the spectrum of ${\hat{H}}$ is now {\em continuous\/} and
consists of the non-negative half-line.

These examples suggest that the continuity versus discreteness of the
spectrum is related to the question of whether the  wormhole throat
reaches the initial and final singularities within finite parameter time.
There are general grounds to expect this to be the case. A~classical
theory in which the two-sphere radius reaches zero within finite parameter
time is singular, in the sense that the classical solutions cannot be
continued arbitrarily far into the past and future. If one quantizes such
a theory so that the classical Hamiltonian is promoted into a
time-independent, self-adjoint Hamiltonian operator, then the unitary
evolution generated by the Hamiltonian operator remains well-defined for
arbitrarily large times. If one starts with an initial wave function that
is a wave packet following some classical trajectory, the quantum time
evolution will force the wave packet to be reflected from the classical
singularity.
The reflection is an entirely quantum mechanical phenomenon, and
the quantum dynamics acquires in this sense a quasiperiodic character.
On the other hand, if the classical
solutions require an infinite amount of time to reach the singularity, one
generically expects \cite{reed-simonII} that in the quantum theory a wave
packet initially following a classical trajectory will just keep following
this trajectory, with some spreading, for arbitrarily large times. It is
clear that these arguments apply without change also in the presence of a
negative cosmological constant. In the presence of a nonvanishing charge,
an analogous discussion applies with the  singularity replaced by the inner
horizon.

One may hold mixed feelings about a wormhole throat quantum theory that
introduces a quantum mechanical bounce at a classical singularity or at an
inner horizon.  On the one hand, singularities and inner horizons are places
where the classical theory behaves poorly, and one anticipates quantum
effects to be important. On the other hand, an outright bounce may
appear an uncomfortably orderly quantum prediction, given that
(semi)classical intuition associates singularities and inner horizons with
collapses and instabilities. Related discussion, in this and related
contexts, can be found in Refs.\
\cite{gotay-dem,poisson-israel,bonanno,marolf-recoll,%
horo-marolf,horo-myers,droz}.
While we view the model of the present paper as a useful arena where these
issues can be addressed in relatively explicit terms, the model is
undoubtedly dynamically too poor to support confident conclusions about the
physical reasonableness of a discrete versus continuous black hole
spectrum. It would be substantially more interesting if our techniques
could be generalized to models containing degrees of freedom that carry
Hawking radiation.

\acknowledgments
We are indebted to John Friedman, Stephen Winters-Hilt, and Ian Redmount
for making us aware of their previous, related work on throat variables for
the Schwarzschild geometry \cite{fried-red-win,redmount-talk}, and to John
Friedman and Stephen Winters-Hilt for discussions and collaboration in an
early stage of this work. We would also like to thank Andrei Barvinsky,
David Brown, Valeri Frolov, Gary Horowitz, Ted Jacobson, Ian Kogan, Karel
Kucha\v{r}, Gabor Kunstatter, Pawel Mazur, Diego Mazzitelli, Yoav
Peleg, Jonathan Simon, Thomas Strobl, and especially Don Marolf, for helpful
discussions and comments.
J.~M. is grateful to the DAMTP relativity group for hospitality during his
visit.
J.~L. was supported in part by the NSF grant PHY91-19726.
J.~M. was supported by the Finnish Cultural Foundation.

\appendix
\section{Semi-boundedness of $\hhzero$}
\label{app:semi-boundedness}

In this appendix we shall show that every self-adjoint extension of the
Hamiltonian $\hhzero$ (\ref{hhzero}) is bounded below, and that certain
self-adjoint extensions are strictly positive. We shall discuss separately
the cases $r\ge 3/2$, $1/2 < r < 3/2$, and $r = 1/2$.

\subsection{$r\ge 3/2$}

For $r\ge 3/2$, $\hhzero$~is essentially self-adjoint. Let $\chi$ be an
eigenfunction with energy~$E$. With a suitable choice of the overall
numerical factor, $\chi$ is real-valued and has the small $x$ expansion
\begin{equation}
\chi(x) = x^r \left( 1 + O(x^2) \right)
\ \ .
\end{equation}
Both $\chi$ and $\chi'$ are therefore positive for sufficiently
small~$x$. If now $E\le0$, the eigenvalue equation shows then that $\chi$
is increasing for all $x>0$. This implies that $\chi$ cannot be
normalizable, which contradicts the assumption that $\chi$ is an
eigenfunction. Hence the spectrum is strictly positive.

\subsection{$1/2 < r < 3/2$}

For $1/2 < r < 3/2$, the self-adjoint extensions of $\hhzero$ form a
family characterized by a boundary condition at $x=0$, and parametrized
by~$U(1)$. To find these extensions, we note that for any~$E$, the
differential equation
$\hhzero\chi=E\chi$ has two linearly independent solutions, denoted by
$u_E(x)$ and $v_E(x)$, with the asymptotic small $x$ behavior
\begin{mathletters}
\label{uv}
\begin{eqnarray}
u_E(x) &=& x^r \left( 1 + O(x^2) \right)
\ \ ,
\\
v_E(x) &=& x^{1-r} \left( 1 + O(x^2) \right)
\ \ .
\end{eqnarray}
\end{mathletters}%
For real~$E$, both $u$ and $v$ are real-valued. Using the techniques of
Ref.\ \cite{reed-simonII}, it is easily shown that the
eigenfunctions of a given self-adjoint extension of $\hhzero$ take, up to
overall normalization, the form
\begin{equation}
\chi_E = \cos(\theta) u_E + \sin(\theta) v_E
\ \ ,
\label{sa}
\end{equation}
where $\theta\in[0,\pi)$ is the parameter specifying the self-adjoint
extension. The condition (\ref{sa}) can be written without explicit
reference to the solutions (\ref{uv}) as
\begin{equation}
0 = \lim_{x\to0}
\left[ (2r-1) \cos(\theta) x^{r-1} \chi - \sin(\theta)
x^{2(1-r)} {d(x^{r-1}
\chi)\over dx}
\right]
\ \ .
\label{sa-formal}
\end{equation}

We now proceed to obtain a lower bound for the eigenenergies.

Consider first an extension in the range $0\le \theta \le \pi/2$. Comparing
the eigenvalue differential equation to the corresponding equation with
the term $\casehalf x^{2/3}$ omitted from $\hhzero$ (\ref{hhzero}) and the
energy set to zero, one sees that the prospective eigenfunctions with
$E\le0$ are bounded below by the function $\cos(\theta) x^r + \sin(\theta)
x^{1-r}$, which does not vanish exponentially at large~$x$. However, as
the potential increases without bound as $x$ goes to infinity, every
eigenfunction must vanish exponentially at large~$x$. Hence the spectrum is
strictly positive.

Consider then an extension in the remaining range $\pi/2 < \theta < \pi$.
Let $\chi$ be an eigenfunction with energy $E<0$. Writing $y =
{(-8E/9)}^{1/2}x$, the eigenfunction equation reads
\begin{equation}
0 = \left[ - {d^2 \over dy^2} + {r(r-1) \over y^2} + 1
+
{\left(3y \over 8 E^2 \right)}^{2/3}
\right] \chi
\ \ .
\label{rescaled-eq}
\end{equation}
The last term in (\ref{rescaled-eq}) is
asymptotically small at large negative~$E$, uniformly in the interval
$y\in(0,M]$, where $M$ is an arbitrary positive constant.
Omitting this last term gives an equation whose linearly independent
solutions are $y^{1/2}I_{r-(1/2)}(y)$ and $y^{1/2}I_{(1/2)-r}(y)$, where
$I$ is a modified Bessel function
\cite{Grad-Rhyz}.
Therefore $\chi$
has at large negative $E$ the asymptotic behavior
\begin{eqnarray}
\chi \sim {(y/2)}^{1/2}
\Bigl[ \,
&&
\cos(\theta) \Gamma(r + \casehalf) {(-2E/9)}^{-r/2} I_{r-(1/2)}(y)
\nonumber
\\
&&+
\sin(\theta) \Gamma(\case{3}{2} - r) {(-2E/9)}^{(r-1)/2} I_{(1/2)-r}(y)
\,
\Bigr]
\ \ ,
\label{chi-as-negE}
\end{eqnarray}
uniformly for $y\in(0,M]$. The coefficients of the two Bessel functions in
(\ref{chi-as-negE}) have been fixed by comparing the small $y$ expansions of
(\ref{sa}) and (\ref{chi-as-negE}) \cite{Grad-Rhyz}.

By the asymptotic behavior of the Bessel functions at large argument
\cite{Grad-Rhyz}, we can now choose $M$ so that $y^{1/2}
I_{(1/2)-r}(y)$ is positive and increasing for $y\ge M/2$. For future use,
we make this choice so that $M>1$. Then, the
second term in (\ref{chi-as-negE}) dominates the first term at large
negative~$E$, uniformly for
$M/2 \leq y \leq M$.  Therefore there exists a constant ${\tilde E}<0$,
dependent on $r$ and~$\theta$, such that
$\chi$ and
$d\chi/dy$ are positive at $y=M$ whenever $E < {\tilde E}$. As
$|r(r-1)| < 1$,  equation (\ref{rescaled-eq}) then shows, by virtue of the
choice $M>1$,  that
$\chi$ diverges at large $y$ whenever $E < {\tilde E}$. As $\chi$ is by
assumption normalizable, we thus see that the eigenenergies are bounded
below by ${\tilde E}$.

Note that the lower bound for the eigenenergies is not uniform
in~$\theta$. For fixed $r$ and any given~$E$, there exists a
unique self-adjoint extension of $\hhzero$ such that $E$ is in the
spectrum. This is because the differential equation $\hhzero\chi=E\chi$
has for any $E$ a normalizable solution that is unique up to
a multiplicative constant, and matching the small $x$ behavior of this
solution to (\ref{sa}) uniquely specifies the value of~$\theta$. One can
thus find extensions with arbitrarily negative ground state energy.

\subsection{$r=1/2$}

For $r=1/2$, the self-adjoint extensions of $\hhzero$ form again a
family parametrized by~$U(1)$. The boundary condition characterizing the
extensions takes the form~(\ref{sa}), where
$\theta\in[0,\pi)$, but now with
\begin{mathletters}
\label{uv-crit}
\begin{eqnarray}
u_E(x) &=& x^{1/2} \left( 1 + O(x^2) \right)
\ \ ,
\\
v_E(x) &=& u_E(x) \ln x \; + O(x^{5/2})
\ \ .
\end{eqnarray}
\end{mathletters}%
Condition (\ref{sa-formal}) is replaced by
\begin{equation}
0 = \lim_{x\to0}
\left\{
\left[ \cos(\theta) + \sin(\theta) \ln x \right]
x {d(x^{-1/2} \chi)\over dx}
- \sin(\theta) x^{-1/2} \chi
\right\}
\ \ .
\label{sa-formal-crit}
\end{equation}

For the extension with $\theta=0$, one sees as above that the spectrum is
strictly positive.

Consider then an extension in the range $0<\theta<\pi$. Let $\chi$ be an
eigenfunction with energy $E<0$, and proceed as above. Equation
(\ref{chi-as-negE}) is replaced by
\begin{equation}
\chi \sim  {\left(-8E\over9\right)}^{-1/4} y^{1/2}
\left\{
\left[ \cos(\theta) - \sin(\theta) \left( \gamma + \casehalf \ln \left(
-2E/9 \right) \right) \right] I_0(y)
- \sin(\theta) K_0(y)  \right\}
\ \ ,
\label{chi-as-negE-crit}
\end{equation}
where $K$ is the second modified Bessel function \cite{Grad-Rhyz} and
$\gamma$ is Euler's constant. At large negative~$E$, the term
proportional to $y^{1/2} I_0(y)$ dominates, and one can argue as above.
Hence the spectrum is bounded below.

As in the case $1/2<r<3/2$, for any given energy $E$ there exists a
self-adjoint extension such that $E$ is in the spectrum. One can thus
again find extensions with arbitrarily negative ground state energy.

\section{Large eigenvalues of $\hhzero$}
\label{app:large-eigen}

In this appendix we analyze the asymptotic distribution of the large
eigenvalues of the self-adjoint extensions of $\hhzero$~(\ref{hhzero}).
The idea is to match a Bessel function approximation at small argument to a
WKB approximation in the region of rapid oscillations. We shall again
discuss separately the cases $r\ge 3/2$, $1/2 < r < 3/2$, and $r = 1/2$.

\subsection{$r\ge 3/2$}

We begin with the case $r\ge 3/2$, where $\hhzero$~is essentially
self-adjoint. We shall throughout denote by $\chi$ an eigenfunction with
energy $E>0$.

Consider first $\chi$ at small argument. Setting $z =
{(8E/9)}^{1/2}x$, the eigenfunction equation reads
\begin{equation}
0 = \left[ - {d^2 \over dz^2} + {r(r-1) \over z^2} - 1
+
{\left(3z \over 8 E^2 \right)}^{2/3}
\right] \chi
\ \ .
\label{rescaled-eq-2}
\end{equation}
The last term in (\ref{rescaled-eq-2}) is
asymptotically small at large~$E$, uniformly in the interval
$z\in(0,ME^{1/2}]$, where $M$ is an arbitrary positive constant.
Omitting this last term gives an equation whose linearly independent
solutions are
$z^{1/2}J_{r-(1/2)}(z)$ and $z^{1/2}N_{r-(1/2)}(z)$, where $J$ and $N$ are
the Bessel functions of the first and second kind, and only the former
solution is normalizable at small~$x$ \cite{Grad-Rhyz}.
The asymptotic
large $E$ behavior of $\chi$ is therefore
\begin{equation}
\chi \aspropto x^{1/2} J_{r-(1/2)}
\left[ {{(8E)}^{1/2}x\over 3} \right]
\ \ ,
\label{chi-aspropto1}
\end{equation}
valid uniformly in any bounded region in~$x$. Here, and from now on, we
use the symbol $\aspropto$ to denote the asymptotic form at large~$E$, up
to a possibly $E$-dependent coefficient. Introducing two constants
$\delta_1$ and $\delta_2$ that satisfy $0< \delta_1 < \delta_2$, and using
the asymptotic large argument behavior of~$J$ \cite{Grad-Rhyz}, we can
rewrite (\ref{chi-aspropto1}) as
\begin{equation}
\chi \aspropto \cos
\left[ {{(8E)}^{1/2}x\over 3} - {\pi r \over 2} \right]
\ \ ,
\label{chi-aspropto2}
\end{equation}
valid uniformly for $\delta_1\le x \le \delta_2$.

Consider then the region of rapid oscillations. We take $E$ so large that
the eigenvalue equation has two turning points, and we denote the
larger turning point by~$x_0$. The region of rapid oscillations at large
$E$ is sufficiently far left of $x_0$ but sufficiently far right of the
origin. As $\chi$ decays exponentially right of~$x_0$, the WKB
approximation to the wave function in the region of rapid oscillations is
(see, for example, Ref.\
\cite{galindoII})
\begin{equation}
\chi_{\rm WKB} = {[p_2(x)]}^{-1/2}
\cos
\left[
\int_x^{x_0} dx' \, p_2(x') - {\pi\over4} \right]
\ \ ,
\label{WKB-chi}
\end{equation}
where
\begin{equation}
p_2 = {2\over3}
{\left[
2E - x^{2/3} - {9 r(r-1) \over 4 x^2}
\right]}^{1/2}
\ \ .
\end{equation}
The evaluation of the integral in (\ref{WKB-chi}) is discussed
in appendix~\ref{app:integral}. We find
\begin{equation}
\chi_{\rm WKB} \aspropto
\cos
\left[
{{(8E)}^{1/2}x\over 3} - {\pi E^2 \over 2} + {\pi \over 4}
+ O \left( E^{-1/2} \right)
\right]
\ \ ,
\label{WKB-chi-aspropto}
\end{equation}
valid uniformly for $\delta_1\le x \le \delta_2$. Comparing
(\ref{chi-aspropto2}) and (\ref{WKB-chi-aspropto}) yields for the large
eigenenergies the WKB estimate
\begin{equation}
E_{\rm WKB}^2 \sim 2k + r + \casehalf + o(1)
\ \ ,
\label{WKB-E1}
\end{equation}
where $k$ is a large integer, and $o(1)$ indicates a term that goes to
zero at large~$E$.

Note that we have not attempted to control how far the integer $k$ is
from the number of the eigenvalue that equation (\ref{WKB-E1}) is meant to
approximate. To obtain a formula that gives an approximation to the $k$th
eigenenergy in the limit of large~$k$, one may need to add to the right
hand side of (\ref{WKB-E1}) some even integer.

As the potential term in the Hamiltonian is smooth without oscillations or
step-like behavior, and as $|p_2' p_2^{-2}|$ vanishes for large $E$
uniformly in $\delta_1\le x \le \delta_2$, one expects the WKB estimate to
the large eigenenergies to be an accurate one. We shall not attempt to
investigate the accuracy rigorously for general~$r$, but we shall see below
that the accuracy can be verified by independent means in the special case
$r=5/6$, and, to leading order, also in the special case $r=1$.

\subsection{$1/2 < r < 3/2$}

For $1/2 < r < 3/2$, we recall from appendix \ref{app:semi-boundedness}
that the self-adjoint extensions of $\hhzero$ are specified by the
boundary condition (\ref{sa}) at small~$x$. Fixing $\theta$ and proceeding
as above, we see
from the small argument behavior of the
Bessel functions \cite{Grad-Rhyz}
that equation (\ref{chi-aspropto1}) is replaced by
\begin{eqnarray}
\chi \sim {(z/2)}^{1/2}
\Bigl[ \,
&&
\cos(\theta) \Gamma(r + \casehalf) {(2E/9)}^{-r/2} J_{r-(1/2)}(z)
\nonumber
\\
&&+
\sin(\theta) \Gamma(\case{3}{2} - r) {(2E/9)}^{(r-1)/2} J_{(1/2)-r}(z)
\,
\Bigr]
\ \ .
\label{chi-as-posE}
\end{eqnarray}
When $\theta=0$, the second term in (\ref{chi-as-posE}) vanishes, and we
can proceed as above, with $x_0$ now denoting the larger one of the
two turning points for $r>1$ and the unique turning point for $r\le1$.
The WKB estimate for the large eigenenergies is again given
by~(\ref{WKB-E1}). When $\theta\ne0$, on the other hand, the second term in
(\ref{chi-as-posE}) dominates the first term at large~$E$, and $r$ in
(\ref{chi-aspropto2}) is replaced by $1-r$.
The WKB estimate (\ref{WKB-E1}) is therefore replaced by
\begin{equation}
E_{\rm WKB}^2 \sim 2k - r - \casehalf + o(1)
\ \ .
\label{WKB-E2}
\end{equation}

In the special case $r=5/6$, we can verify the accuracy of these WKB
results rigorously. The eigenfunctions are now $\chi =
x^{1/6}U \! \left(-
\casehalf E^2,
\sqrt{2} ( x^{2/3} - E) \right)$, where
$U$ is the parabolic cylinder function that vanishes at large values of its
second argument \cite{Ab-Steg}. The boundary condition (\ref{sa-formal})
reads
\begin{equation}
0 =
\cos(\theta)
U \! \left(- \casehalf E^2, - \sqrt{2} E
\right)
- \sqrt{2} \sin(\theta)
U' \! \left(- \casehalf E^2, - \sqrt{2} E
\right)
\ \ ,
\label{parabolic-sa}
\end{equation}
where the prime denotes the derivative of $U$ with respect to its second
argument. Using Olver's asymptotic expansions of parabolic cylinder
functions (Ref.\ \cite{olver}, formula~(9.7), and the discussion of the
derivative on p.~155), we find
\begin{mathletters}
\begin{eqnarray}
E^2 &\sim& 2k - 2/3 + O(E^{-8/3})
\qquad \hbox{for $\theta=0$}
\ \ ,
\\
E^2 &\sim& 2k + 2/3 + O(E^{-4/3})
\qquad \hbox{for $\theta=\pi/2$}
\ \ ,
\\
E^2 &\sim& 2k + 2/3 + O(E^{-1/3})
\qquad \hbox{for $0\neq\theta\neq\pi/2$}
\ \ .
\end{eqnarray}
\end{mathletters}%
This corroborates the WKB results (\ref{WKB-E1}) and (\ref{WKB-E2})
for $r=5/6$, and gives an improved bound for the error term.

In the special case $r=1$, the theorem in \S7 of Ref.\ \cite{atkinson}
yields the rigorous asymptotic estimate
\begin{equation}
E^2 \sim 2k + O(\ln E)
\ \ .
\label{atkinson-result}
\end{equation}
(The leading order term of (\ref{atkinson-result}) also follows from
Ref.\ \cite{dunfordII}, p.~1614.) This corroborates the leading order term
in our WKB results (\ref{WKB-E1}) and (\ref{WKB-E2}) for $r=1$.

Finally, we note in passing that for $r=5/6$, the boundary condition
(\ref{parabolic-sa}) and Olver's expansions
\cite{olver} yield the asymptotic relation
\begin{equation}
\tan(\theta) \sim
-
{3^{1/6} {\left[ \Gamma(1/3) \right]}^2
\over
2^{4/3} \pi {(-E_0)}^{1/3} }
+ O \left( {(-E_0)}^{-5/3} \right)
\end{equation}
for the parameter $\theta$ and the ground state energy~$E_0$, valid in the
limit of large negative~$E_0$.

\subsection{$r=1/2$}

For $r=1/2$, equation (\ref{chi-as-posE}) is replaced by
\begin{equation}
\chi \sim  {\left(8E\over9\right)}^{-1/4} z^{1/2}
\left\{
\left[ \cos(\theta) - \sin(\theta) \left( \gamma + \casehalf \ln
\left( 2E/9 \right) \right) \right] J_0(z)
+ (\pi/2) \sin(\theta) N_0(z)  \right\}
\ \ ,
\label{chi-as-posE-crit}
\end{equation}
where $\gamma$ is Euler's constant as before.
For any~$\theta$, the term proportional to $z^{1/2} J_0(z)$ dominates at
large energies, and equation (\ref{chi-aspropto2}) holds with $r= 1/2$. The
WKB estimate for the eigenenergies is thus given by (\ref{WKB-E1}) with $r=
1/2$, for any~$\theta$.

\section{Evaluation of the WKB integral}
\label{app:integral}

In this appendix we outline the evaluation of the integral in
(\ref{WKB-chi}) at large~$E$. We shall throughout assume that $E$ is so
large that a classically allowed domain exists (which is a restriction
only for $r>1$), that $x$ is in the classically allowed domain, and that
$x>\delta_1$, where the constant $\delta_1$ was introduced in
appendix~\ref{app:large-eigen}.

Returning to the variable $a=x^{2/3}$ of the main text, and writing
$b={(x')}^{2/3}$, the integral in the exponent in (\ref{WKB-chi}) takes the
form
\begin{equation}
S :=
\int_a^{a_0} db
\sqrt{2Eb - b^2}
\sqrt{
1 - {9r(r-1) \over 4 b^3 (2E - b)}
}
\ \ ,
\label{int:S}
\end{equation}
where $a_0=x_0^{2/3}$ is the (larger) turning point. At large~$E$, $a_0
\sim 2E + O( E^{-3})$.

We fix a constant $\delta_3>0$, take $E$ so large that $|a_0 - 2E| <
\delta_3/E$, and restrict $a$ to be in the interval $\delta_1^{2/3}
\le a \le 2E - \delta_3/E$. We can then replace the upper limit
of the integral in (\ref{int:S}) by $2E - \delta_3/E$, with the error in
$S$ being of order~$O( E^{-1})$. The second term under the second
square root in (\ref{int:S}) is now uniformly of order~$O( E^{-1})$. As the
first square root is of order~$O(E)$, we can expand the second square root
in the Taylor series and truncate the series after the third term, with the
error in $S$ being of order~$O( E^{-1})$. In the
truncated integrand, the third term is proportional to $b^{-11/2}
{(2E-b)}^{-3/2}$, which is uniformly of order~$O( E^{-3/2})$, and this term
can thus be omitted with the consequence of making an error of order $O(
E^{-1/2})$ in~$S$. We therefore have
\begin{equation}
S \sim \int_a^{2E - \delta_3/E}
db
\left[
\sqrt{2Eb - b^2}
- {9r(r-1) \over 8 b^2 \sqrt{2Eb - b^2} }
\right]
\ \ + O(E^{-1/2})
\ \ .
\label{int:S2}
\end{equation}
The integral in (\ref{int:S2}) is elementary. The contribution from the
second term turns out to be of order~$O(E^{-1/2})$, and in the first term
the upper limit can be replaced by $2E$ with the consequence of making an
error of order~$O( E^{-1})$. We thus obtain
\begin{equation}
S \sim \casehalf (E-a) \sqrt{2Ea - a^2}
+ \casehalf E^2 \arcsin\left(1 - a / E\right)
+ \case{1}{4}
\pi E^2
+ O(E^{-1/2})
\ \ .
\label{int:S3}
\end{equation}

Finally, we restrict $a$ to be in the interval $\delta_1^{2/3} \le a \le
\delta_2^{2/3}$, where the constant $\delta_2$ was introduced
in appendix~\ref{app:large-eigen}. A~large $E$ expansion of (\ref{int:S3})
then gives
\begin{equation}
S \sim
- {\sqrt{8Ea^3} \over 3}
+ {\pi E^2 \over 2 }
+ O(E^{-1/2})
\ \ ,
\end{equation}
which leads to~(\ref{WKB-chi-aspropto}).


\newpage

\begin{references}

\bibitem{cartei1}
S.~Carlip and C.~Teitelboim,
Class.\ Quantum Grav.\ {\bf 12}, 1699 (1995).
(gr-qc/9312002)

\bibitem{cartei2}
S.~Carlip and C.~Teitelboim,
Phys.\ Rev.\ D {\bf 51}, 622 (1995).
(gr-qc/9405070)

\bibitem{carlip}
S.~Carlip,
Phys.\ Rev.\ D {\bf 51}, 632 (1995).
(gr-qc/9409052)

\bibitem{bala-edge}
A.~P. Balachandran, L.~Chandar, and A.~Momen,
Nucl.\ Phys.\ {\bf B461}, 581 (1996).
(gr-qc/9412019)

\bibitem{bombelli+}
L.~Bombelli,
R.~Koul,
J.~Lee, and
R.~D.~Sorkin,
Phys.\ Rev.\ D {\bf 34}, 373 (1986).

\bibitem{srednicki}
M.~Srednicki,
Phys.\ Rev.\ Lett.\ {\bf 71}, 666 (1993).

\bibitem{fro-nov}
V.~Frolov and
I.~Novikov,
Phys.\ Rev.\ D {\bf 48}, 4545 (1993).
(gr-qc/9309001)

\bibitem{frolov}
V.~P. Frolov,
Phys.\ Rev.\ Lett.\ {\bf 74}, 3319 (1995).
(gr-qc/9406037)

\bibitem{horo-rev}
G.~T. Horowitz,
``The origin of black hole entropy in string theory,"
Report UCSBTH-96-07, gr-qc/9604051, to appear in the proceedings of the
Pacific Conference on Gravitation and Cosmology, Seoul, Korea, February
1-6, 1996.

\bibitem{page-inf-rev}
D.~N. Page,
in {\it Proceedings of the
5th Canadian Conference on General Relativity and Relativistic
Astrophysics,
University of Waterloo, 13--15 May, 1993},
edited by R.~B. Mann and
R.~G. McLenaghan (World Scientific, Singapore, 1994).
(hep-th/9305040)

\bibitem{beken-rev}
J.~D. Bekenstein,
in
{\it General Relativity\/}, Proceedings of the Seventh Marcel Grossman
Meeting, Stanford, California, 1994, edited by R.~Ruffini and M.~Keiser
(Word Scientific, Singapore, 1995).
(gr-qc/9409015)

\bibitem{hawkingCMP}
S.~W. Hawking,
Commun.\ Math.\ Phys.\ {\bf 43}, 199 (1975).

\bibitem{bekenstein1}
J.~D. Bekenstein,
Lett.\ Nuovo Cimento {\bf 11}, 467 (1974).

\bibitem{mukha1}
V.~F. Mukhanov,
Pis'ma Zh.\ Eksp.\ Teor.\ Fiz.\ {\bf 44}, 50 (1986)
[JETP Lett.\ {\bf 44}, 63 (1986)].

\bibitem{kogan1}
I.~Kogan,
Pis'ma Zh.\ Eksp.\ Teor.\ Fiz.\ {\bf 44}, 209 (1986)
[JETP Lett.\ {\bf 44}, 267 (1986)].

\bibitem{mazur-grg}
P.~O. Mazur,
Gen.\ Relativ. Gravit.\ {\bf 19}, 1173 (1987). 

\bibitem{mazur-string}
P.~O. Mazur,
Phys.\ Rev.\ Lett.\ {\bf 57}, 929 (1986);
{\bf 59}, 2380 (1987).

\bibitem{mukha2}
V.~F. Mukhanov,
in {\it Complexity, Entropy, and the Physics of
Information\/},
SFI Studies in the Sciences of Complexity, Vol.\ III,
edited by W.~H. Zurek
(Addison-Wesley, New York, 1990).

\bibitem{danielsson}
U.~H. Danielsson and
M.~Schiffer,
Phys.\ Rev.\ D {\bf 48}, 4779 (1993).

\bibitem{bellido}
J.~Garc\'{\i}a--Bellido,
``Quantum Black Holes,"
Report SU-ITP-93/4, hep-th/9302127.

\bibitem{peleg1}
Y.~Peleg,
``Quantum dust black holes,"
Report BRX-TH-350, hep-th/9307057.

\bibitem{maggiore}
M.~Maggiore,
Nucl.\ Phys.\ {\bf B429}, 205 (1994).
(gr-qc/9401027)

\bibitem{kogan2}
I.~Kogan,
``Black Hole Spectrum, Horison Quantization and All That,"
Report OUTP--94--39P, hep-th/9412232.

\bibitem{lousto2}
C.~O. Lousto,
Phys.\ Rev.\ D {\bf 51}, 1733 (1995).
(gr-qc/9405048)

\bibitem{peleg2}
Y.~Peleg,
Phys.\ Lett.\ B {\bf 356}, 462 (1995).

\bibitem{bek-mu}
J.~D. Bekenstein and
V.~F. Mukhanov,
Phys.\ Lett.\ B {\bf 360}, 7 (1995).
(gr-qc/9505012)

\bibitem{mazur-pol1}
P.~O. Mazur,
Acta Phys.\ Pol.\  B {\bf 26}, 1685 (1995).
(hep-th/9602044)

\bibitem{mazur-pol2}
P.~O. Mazur,
``Gravitation, the Quantum, and Cosmological Constant,"
e-print hep-th/9603014.

\bibitem{kastrup}
H.~A. Kastrup,
``On the quantum levels of isolated spherically symmetric gravitational
systems," Report PITHA 96/16, gr-qc/9605038.

\bibitem{barvi-kunst1}
A.~Barvinsky and
G.~Kunstatter,
``Exact Physical Black Hole States in Generic 2-D Dilaton Gravity,"
e-print hep-th/9606134.

\bibitem{barvi-kunst2}
A.~Barvinsky and
G.~Kunstatter,
``Mass Spectrum for Black Holes in Generic 2-D Dilaton Gravity,"
e-print gr-qc/9607030.

\bibitem{MTW}
C.~W. Misner, K.~S. Thorne, and J.~A. Wheeler,
{\it Gravitation\/} (Freeman, San Francisco, 1973).

\bibitem{lousto3}
C.~O. Lousto,
Phys.\ Lett.\ B {\bf 352}, 228 (1995).
(gr-qc/9411037)

\bibitem{barreira}
M.~Barreira,
M.~Carfora,  and
C.~Rovelli,
Gen.\ Relativ.\ Gravit., to be published.
(gr-qc/9603064).

\bibitem{exact-book}
D.~Kramer, H.~Stephani, E.~Herlt, and
M.~MacCallum,
{\it Exact Solutions of Einstein's Field Equations\/},
edited by E.~Schmutzer (Cambridge University Press, Cambridge,
England, 1980), Sec.\ 13.4.

\bibitem{thiemann1}
T.~Thiemann and H.~A. Kastrup,
Nucl.\ Phys.\ {\bf B399}, 211 (1993).
(gr-qc/9310012)

\bibitem{thiemann2}
H.~A. Kastrup and T.~Thiemann,
Nucl.\ Phys.\ {\bf B425}, 665 (1994).
(gr-qc/9401032)

\bibitem{kuchar1}
K.~V. Kucha\v{r},
Phys.\ Rev.\ D {\bf 50}, 3961 (1994).
(gr-qc/9403003)

\bibitem{LW2}
J.~Louko and
B.~F. Whiting,
Phys.\ Rev.\ D {\bf 51}, 5583 (1995).
(gr-qc/9411017)

\bibitem{dunfordII}
N.~Dunford and J.~S. Schwartz,
{\it Linear Operators\/}
(Interscience, New York, 1963),
Vol.~II.

\bibitem{reed-simonII}
M.~Reed and B.~Simon,
{\it Methods of Modern Mathematical Physics\/}
(Academic, New York, 1975),
Vol.~II.

\bibitem{lake1}
K.~Lake,
Phys.\ Rev.\ D {\bf 19}, 421 (1979).

\bibitem{btz-cont}
M.~Ba\~{n}ados,
C.~Teitelboim,
and
J.~Zanelli,
Phys.\ Rev.\ D {\bf 49}, 975 (1994).
(gr-qc/9307033)

\bibitem{thiemann3}
T.~Thiemann,
Int.\ J.\ Mod.\ Phys.\ D {\bf 3}, 293 (1994).

\bibitem{thiemann4}
T.~Thiemann,
Nucl.\ Phys.\ {\bf B436}, 681 (1995).

\bibitem{lou-win}
J.~Louko and S.~N. Winters-Hilt,
Phys.\ Rev.\ D {\bf 54}, 2647 (1996).
(gr-qc/9602003)

\bibitem{haw-ell}
S.~W. Hawking and
G.~F.~R. Ellis,
{\it The Large Scale
Structure of Space-time\/}
(Cambridge University Press, Cambridge, England, 1973).

\bibitem{arfken}
G.~Arfken,
{\it Mathematical Methods for Physicists\/}, 2nd edition
(Academic, New York, 1970).

\bibitem{woodhouse}
N.~M.~J. Woodhouse,
{\it Geometric Quantization\/}
(Clarendon Press, Oxford, 1980).

\bibitem{isham-LH}
C. J.~Isham,
in {\it Relativity, Groups and Topology II:
Les Houches 1983,}
edited by B.~S. DeWitt and R.~Stora
(North-Holland, Amsterdam, 1984).

\bibitem{AAbook}
A.~Ashtekar,
{\it Lectures on Non-Perturbative Canonical Gravity\/}
(World Scientific, Singapore, 1991).

\bibitem{cava1}
M.~Cavagli\`a, V. de~Alfaro, and A.~T. Filippov,
Int.\ J.\ Mod.\ Phys.\ D {\bf 4}, 661 (1995).
(gr-qc/9411070)

\bibitem{cava2}
M.~Cavagli\`a, V. de~Alfaro, and A.~T. Filippov,
``Quantization of the Schwarzschild Black Hole,"
Report DFTT 50/95, gr-qc/9508062.

\bibitem{topocen}
J.~L. Friedman, K.~Schleich, and D.~M. Witt,
Phys.\ Rev.\ Lett.\ {\bf 71}, 1486 (1993).

\bibitem{Ab-Steg}
{\it Handbook of Mathematical Functions,\/}
edited by M.~Abramowitz and I.~A. Stegun
(Dover, New York, 1965).

\bibitem{gotay-dem}
M.~J. Gotay and J.~Demaret,
Phys.\ Rev.\ D {\bf 28}, 2402 (1983).

\bibitem{poisson-israel}
E.~Poisson and
W.~Israel,
Phys.\ Rev.\ D {\bf 41}, 1796 (1990).

\bibitem{bonanno}
A.~Bonanno,
S.~Droz,
W.~Israel, and
S.~M. Morsink,
Phys.\ Rev.\ D {\bf 50}, 7372 (1994).
(gr-qc/9403019)

\bibitem{marolf-recoll}
D.~Marolf,
Class.\ Quantum Grav.\ {\bf 12}, 1199 (1995).
(gr-qc/9404053)

\bibitem{horo-marolf}
G.~T. Horowitz and
D.~Marolf,
Phys.\ Rev.\ D {\bf 52}, 5670 (1995).
(gr-qc/9504028)

\bibitem{horo-myers}
G.~T. Horowitz and
R.~Myers,
Gen.\ Relativ.\ Gravit.\ {\bf 27}, 915 (1995).
(gr-qc/9503062)

\bibitem{droz}
S.~Droz,
W.~Israel, and
S.~M. Morsink,
{\it Phys.\ World\/}
{\bf 1996}(1), 34 (1996).

\bibitem{fried-red-win}
J.~L. Friedman, I.~H. Redmount, and S.~N. Winters-Hilt
(unpublished).

\bibitem{redmount-talk}
I.~H. Redmount, talk given at the 3rd Midwest Relativity Meeting,
Oakland, Michigan, 1993.

\bibitem{Grad-Rhyz}
I.~S. Gradshteyn and
I.~M. Ryzhik,
{\it Table of Integrals, Series and Products,}
4th edition (Academic Press, New York, 1980).

\bibitem{galindoII}
A. Galindo and
P. Pascual,
{\it Quantum Mechanics}
(Springer, Berlin, 1991),
Vol.~II.

\bibitem{olver}
F.~W.~J. Olver,
J. Res.\ Nat.\ Bur.\ Stand.\  {\bf 63B}, 131 (1959).

\bibitem{atkinson}
F.~V. Atkinson,
Univ.\ Nac.\ Tucum\'an.\ Revista A {\bf 8}, 71 (1951).


\end{references}
\end{document}